\documentstyle[12pt]{article}
\input epsfig.sty

\font\blackboard=msbm10 at 12pt
\font\blackboards=msbm7
\font\blackboardss=msbm5
\newfam\black
\textfont\black=\blackboard
\scriptfont\black=\blackboards
\scriptscriptfont\black=\blackboardss

\newcommand{\junk}[1]{}

\newcommand{\ba}{\begin{array}}
\newcommand{\ea}{\end{array}}
\newcommand{\be}{\begin{equation}}
\newcommand{\ee}{\end{equation}}
\newcommand{\bea}{\begin{eqnarray}}
\newcommand{\eea}{\end{eqnarray}}
\newcommand{\beas}{\begin{eqnarray*}}
\newcommand{\eeas}{\end{eqnarray*}}

\def\laplace{{\kern1pt\vbox{\hrule height 1.2pt\hbox{\vrule width
1.2pt\hskip
  3pt\vbox{\vskip 6pt}\hskip 3pt\vrule width 0.6pt}\hrule height
  0.6pt}
  \kern1pt}}
\def\scriptlap{{\kern1pt\vbox{\hrule height 0.8pt\hbox{\vrule width
  0.8pt
  \hskip2pt\vbox{\vskip 4pt}\hskip 2pt\vrule width 0.4pt}\hrule height
  0.4pt}
  \kern1pt}}

\def\roughly#1{\raise.3ex\hbox{$#1$\kern-
.75em\lower1ex\hbox{$\sim$}}}

\newcommand{\NP}{{\em Nucl.\ Phys.\ }}

\newcommand{\PL}{{\em Phys.\ Lett.\ }}

\newcommand{\CMP}{{\em Comm.\ Math.\ Phys.\ }}
\newcommand{\MPL}{{\em Mod.\ Phys.\ Lett.\ }}

\newcommand{\gone}[1]{}
\begin{document}
\pagestyle{plain}
\setcounter{page}{1}

\baselineskip16pt

\begin{titlepage}

\begin{flushright}
MIT-CTP-2942\\
hep-th/0001201
\end{flushright}
\vspace{13 mm}

\begin{center}

{\Large \bf D-brane effective field theory from string field theory
\\}

\end{center}

\vspace{7 mm}

\begin{center}

Washington Taylor

\vspace{3mm}
{\small \sl Center for Theoretical Physics} \\
{\small \sl MIT, Bldg.  6-306} \\
{\small \sl Cambridge, MA 02139, U.S.A.} \\
{\small \tt wati@mit.edu}\\
\end{center}

\vspace{8 mm}

\begin{abstract}
Open string field theory is considered as a tool for deriving the
effective action for the massless or tachyonic fields living on
D-branes.  Some simple calculations are performed in open bosonic
string field theory which validate this approach.  The level
truncation method is used to calculate successive approximations to
the quartic terms $\phi^4$, $(A^\mu A_\mu)^2$ and $[A_\mu, A_\nu]^2$
for the zero momentum tachyon and gauge field on one or many bosonic
D-branes.  We find that the level truncation method converges for
these terms within 2-4\% when all massive fields up to level 20 are
integrated out, although the convergence is slower than exponential.
We discuss the possibility of extending this work to determine the
structure of the nonabelian Born-Infeld theory describing the gauge
field on a system of many parallel bosonic or supersymmetric D-branes.
We also describe a brane configuration in which tachyon condensation
arises in both the gauge theory and string field theory pictures.
This provides a natural connection between recent work of Sen and
Zwiebach on tachyon condensation in string field theory and unstable
vacua in super Yang-Mills and Born-Infeld field theory.
\end{abstract}

\vspace{1cm}
\begin{flushleft}
January 2000
\end{flushleft}
\end{titlepage}
\newpage

\section{Introduction}

String field theory provides an off-shell formulation of string theory
which has the potential to address nonperturbative questions in a
systematic fashion.  A particularly elegant formulation of covariant
string field theory for an open bosonic string was provided by Witten
\cite{Witten-SFT}.  In this string field theory, the entire classical
action is contained in a pair of terms which are quadratic and cubic
in the string field.  The simplicity of this theory makes it feasible
to perform interesting off-shell calculations.  

The bosonic open string has a tachyonic instability around the usual
string vacuum.  Early work by Kostelecky and Samuel \cite{ks-open} and
by Kostelecky and Potting \cite{Kostelecky-Potting} indicated that
string field theory could be used to describe the condensation of the
tachyon, leading to another more stable vacuum\footnote{Another
approach to describing the stable vacuum was taken in
\cite{Bardakci-tachyon}}.  Since, however, the structure of the vacuum
and the role of the tachyonic instability were poorly understood at
that time, the significance of this result was not generally
appreciated.  Recently, renewed interest in tachyonic instabilities
related to D-branes led Sen to propose that the condensation of a
tachyon in the open bosonic string should be understood as the decay
of an unstable D-brane in the bosonic string theory.  The vacuum
energy of the open bosonic string field theory in the ``true'' vacuum
should thus differ from that in the unstable vacuum by precisely the
mass energy of the unstable D-brane, and Sen argued that this mass
difference should be calculable using the bosonic open string field
theory \cite{Sen-tachyon}.  This conclusion was given very strong
support by a recent paper by Sen and Zwiebach \cite{Sen-Zwiebach}, in
which the authors demonstrated that when the string field theory is
truncated to states of level 4, the energy at the minimum of the
potential differs from the energy in the unstable vacuum by 99\% of
the energy of the bosonic D-brane.

The result of Sen and Zwiebach and the earlier work of Kostelecky,
Samuel and Potting suggest that string field theory may
be a much more effective tool for asking nonperturbative and off-shell
questions about string theory than was previously suspected.  While
string field theory has been studied for quite some time, there are
still many fundamental questions about the structure of these
theories.  Even in the context of the simplest string field theory,
that of the open bosonic string, the convergence properties of the
theory are poorly understood.  The work in this paper provides new
evidence that string field theory provides a systematic and completely
convergent framework in which to study many features of string theory.

The main goal of this paper is to initiate a systematic study of how
effective field theories on D-branes can be derived from open string
field theory.  We focus on some of the simplest terms in the effective
actions for the tachyon and massless vector field in the open bosonic
string, namely the quartic terms $\phi^2$ and $(A^\mu A_\mu)^2$ in the
abelian theory and $[A_\mu, A_\nu]^2$ in the nonabelian theory.  We
perform a systematic truncation of string field theory, including all
fields up to a fixed level $n$ for various values of $n \leq 20$.  We
find that the contributions to each of the quartic terms from fields
at level $n$ decrease monotonically and appear to give completely
convergent series with slower than exponential convergence.

These results increase significantly our confidence in the viability
of string field theory as a tool for calculating higher order terms in
the action for the massless/tachyonic fields on the D-brane.  We
discuss briefly how these results may be extended to study the
structure of the nonabelian Born-Infeld theory, and we suggest that
further insight into the process of tachyon condensation may be
gleaned from consideration of simple D-brane models in which the
tachyon, the unstable vacuum and the stable vacuum can all be described
in the language of supersymmetric Yang-Mills theory.

Except for the discussion of tachyon condensation in the last section,
this paper primarily focuses on the open bosonic string field theory.
It would obviously be of great interest to develop an analogous
systematic approach to computations in supersymmetric open
\cite{Witten-SFT-2} and closed \cite{Zwiebach-SFT} string field
theory.  While Witten's formulation of open string field theory can be
extended to the supersymmetric theory
\cite{Witten-SFT-2,Gross-Jevicki-3}, this formalism may be problematic
due to the necessity of considering higher order contact interactions
\cite{Wendt}.  Several alternative formulations of open superstring
field theory have been suggested \cite{pty,Berkovits-general}.  Recent
work of Berkovits \cite{Berkovits-tachyon}, in which he finds a
similar result to that of Sen and Zwiebach for tachyon condensation in
a supersymmetric theory, indicates that his alternative formulation of
the supersymmetric open string field theory \cite{Berkovits-general}
may present a viable framework for systematic calculations.

In Section 2 we review the basic structure of open bosonic string
field theory and describe the calculation of the quartic terms in the
tachyon and vector field potential at zero momentum.  Some
details of the calculations, which were carried out using the symbolic
manipulation program {\it Mathematica}, are described in the
appendices.  In section 3 we discuss the possible extension of this
work to studying the nonabelian Born-Infeld theory and tachyon
condensation in D-brane systems.

\section{Open bosonic string field theory}

In this section we perform some simple examples of calculations in the
open bosonic string field theory to demonstrate the effectiveness of
this approach in finding terms in the effective action for massless
or tachyonic fields on a D-brane.  We begin in subsection
\ref{sec:review} with a brief review of bosonic open string field
theory and the Fock space representation of the Witten vertex, in
order to fix conventions and notation.  In subsection
\ref{sec:tachyon} we calculate the terms in the effective action which
are quartic in the tachyon by integrating out all non-tachyonic
fields.  In subsection \ref{sec:gauge} we calculate the terms quartic
in the gauge field by integrating out the tachyon and all massive
fields.  For all these calculations we describe the first few terms
analytically and summarize the results of the summation of higher
level terms.  Some details of the exact calculations, which were
performed using the symbolic manipulation program {\it Mathematica}
are given in the appendices.

\subsection{Review of notation and formalism}
\label{sec:review}

As formulated by Witten \cite{Witten-SFT}, covariant open string field
theory is described in terms of a string field $\Phi$ which contains a
component field for every state in the first-quantized string Fock
space.  The string field theory action is
\begin{equation}
S = \frac{1}{2 \alpha'}  \int \Phi \star Q \Phi + \frac{g}{3!}  \int \Phi \star
\Phi \star \Phi,
\label{eq:SFT-action}
\end{equation}
where $Q$ is the BRST operator and $\star$ is the string field theory star
product.

Throughout this paper we use the conventions of Kostelecky and Samuel
from \cite{ks-open}; for a more detailed review of open string field
theory see \cite{lpp,Gaberdiel-Zwiebach}.  We summarize here briefly
the basic formalism we will need, following
\cite{Gross-Jevicki-12,cst,Samuel,ks-open}.  The open bosonic string
has a system of 26 matter oscillators $\alpha^{\mu}_{n}$ and ghost
oscillators $b_n, c_n$, which act on the vacuum $| 0 \rangle$ through
\begin{eqnarray*}
\alpha_n^\mu| 0 \rangle & = &  0, \;\;\;\;\;n \geq 0\\
b_n| 0 \rangle & = &  0, \;\;\;\;\;n \geq 0\\
c_n| 0 \rangle & = &  0, \;\;\;\;\;n > 0\\
{}[\alpha_n^\mu, \alpha_m^\nu] & = &n \eta^{\mu \nu} \delta_{n + m} \\
\{b_n, c_m\} & = &  \delta_{n + m}
\end{eqnarray*}
The vacuum $| 0 \rangle$ is related to the SL(2,R) invariant vacuum $|
\Omega \rangle$ through 
\[
| 0 \rangle= c_1| \Omega \rangle.
\]
The string field $\Phi$ is restricted to a sum over states with ghost
number 1 in the classical action.  This restricts us to states in
which an equal number of $b$ and $c$ raising operators act on the
vacuum $| 0 \rangle$.  We denote the Fock space of physical states by
${\cal H}$ and its dual space by ${\cal H}^*$.  We will work in
Feynman-Siegel gauge, where we can impose a further restriction to
states $| S \rangle$ satisfying
\[
b_0 | S \rangle  = 0
\]
The level of a state is defined to be the sum of the oscillator
numbers $n$ used to produce the state from the vacuum $| 0 \rangle$;
thus, the vacuum is the unique state at level 0, $\alpha^\mu_{-1}| 0
\rangle$ are the allowed states at level 1  (of ghost number 1), etc.
These are the states associated with the tachyon and gauge field on
the D-brane.  In our calculations, we will only be interested in
fields which couple through the cubic string field vertex to two
tachyons or two gauge fields.  Thus, we need only consider massive
fields which have both an even level number and an even number of
space-time indices.  In the Feynman-Siegel gauge, the leading terms of
interest in an explicit expansion of the string field $\Phi$ are
\begin{equation}
\Phi = \left( \phi + A_\mu \alpha^\mu_{-1}
 + \frac{1}{\sqrt{2}} B_{\mu \nu} \alpha^\mu_{-1} \alpha^\nu_{-1} +
 \beta b_{-1} c_{-1}  + \cdots \right)| 0 \rangle
\label{eq:expansion}
\end{equation}

It is possible to write both the quadratic kinetic term and the cubic
interaction term from the string field action (\ref{eq:SFT-action}) in
terms of the Fock space representation of the string field.  The
quadratic term evaluated on a state $| S \rangle$ is simply
\begin{equation}
\int | S \rangle \star Q | S \rangle =
\langle S | c_0 \left(\alpha' p^2 +\frac{1}{2} M^2 \right)| S \rangle
\label{eq:kinetic}
\end{equation}
where $\frac{1}{2}M^2$ is the string mass operator, given by the level
of $S$ minus 1, $p$ is the momentum of the state $S$, $\langle S |$ is
the BPZ dual state to $| S \rangle$ produced by acting with the
conformal transformation $z \rightarrow -1/z$, and the dual
vacuum satisfies $\langle 0 | c_0 | 0 \rangle = 1$.

From (\ref{eq:kinetic}) we can write the kinetic terms in the string
field action
\begin{eqnarray}
S_2 & = &  \frac{1}{2 \alpha'}  \int \Phi \star Q \Phi  \nonumber\\
 & = & \frac{1}{2}\int d^{26} x\;\left[
\frac{1}{\alpha'}  \left( -\phi^2 + B_{\mu \nu} B^{\mu \nu} -\beta^2
+ \cdots
 \right)
\right.\label{eq:s2}\\
& &\hspace{0.8in}  \left.
+ \partial_\mu \phi \,\partial^\mu \phi 
+ \partial_\mu A_\nu \,\partial^\mu A^\nu 
+ \partial_\mu B_{\nu \lambda} \, \partial^\mu B^{\nu \lambda}
- \partial_\mu \beta \, \partial^\mu \beta + \cdots \right] \nonumber
\end{eqnarray}
for the terms of interest up to level 2.

The cubic interaction terms in the string field action can be
described in terms of Witten's vertex operator $V$ through
\begin{equation}
\int \Phi \star\Phi\star\Phi =
\langle V | (\Phi \otimes \Phi \otimes \Phi)
\end{equation}
where $V$ is a state in the tensor product space ${\cal H}^*\otimes
{\cal H}^*\otimes {\cal H}^*$.  An explicit representation of
this state in the string Fock space was found in
\cite{Gross-Jevicki-12,cst,Samuel} and can be written as
\begin{equation}
\langle V | = 
\delta (p_{(1)} +p_{(2)} + p_{(3)})
 (\langle 0 | c_0^{(1)} \otimes \langle 0 |c_0^{(2)}
 \otimes\langle 0 | c_0^{(3)})
\exp \left(\frac{1}{2} \alpha^{(r) \mu}_n N^{rs}_{nm} \eta_{\mu \nu} 
 \alpha^{(s)\nu}_{m} + c^{(r)}_n X^{rs}_{nm} b^{(s)}_m \right).
\label{eq:interactions}
\end{equation}
The indices $r, s$ take values from 1-3 and indicate which Fock space
the oscillators act in.  The coefficients $N^{rs}_{nm},X^{rs}_{nm}$
are given in terms of the 6-string Neumann functions
$\bar{N}^{rs}_{nm}, 1 \leq r, s \leq 6$ through
\begin{eqnarray}
N^{rs}_{nm} & = &  \frac{1}{2}
 (\bar{N}^{rs}_{nm} +\bar{N}^{r(s+3)}_{nm}
+\bar{N}^{(r + 3)s}_{nm}+\bar{N}^{(r + 3)(s+3)}_{nm}) \nonumber\\
X^{rs}_{nm} & = &  -m (\bar{N}^{rs}_{nm}-\bar{N}^{r(s+3)}_{nm}),
 \;\;\;\;\; s  = r, r + 2 \label{eq:nx}\\
X^{rs}_{nm} & = &  m (\bar{N}^{rs}_{nm}-\bar{N}^{r(s+3)}_{nm}),
 \;\;\;\;\; s  = r + 1 \nonumber
\end{eqnarray}
For $s < r$ the values of $X^{rs}_{nm}$ are fixed by using
(\ref{eq:nx}) and the cyclic symmetry of the coefficients under
$r \rightarrow
(r\;{\rm mod} \; 3) + 1 $, $s \rightarrow (s\;{\rm mod} \; 3) + 1 $.
When all momenta are zero we only need the Neumann functions with $n,
m > 0$, which are given by
\begin{equation}
\bar{N}^{rs}_{nm} = \frac{1}{nm}  
\oint_{z^r} \frac{dz}{2 \pi i} 
\oint_{z^s} \frac{dw}{2 \pi i} 
\frac{1}{ (z-w)^2} 
(-1)^{n (r-1) + m (s-1)}
(f (z))^{(-1)^r n} (f (w))^{(-1)^s m}
 \label{eq:n}
\end{equation}
with
\begin{equation}
f (z) = \frac{z (z^2 -3)}{3z^2 -1} 
\end{equation}
and
\begin{equation}
z^1, \ldots, z^6 = \sqrt{3}, 1/\sqrt{3},0, -1/\sqrt{3}, -\sqrt{3}, \infty.
\end{equation}
A table of the coefficients $N^{rs}_{nm},X^{rs}_{nm}$ with $n, m < 9$
is given in Appendix A.

From (\ref{eq:interactions}) the cubic interaction terms in the string
field theory can be written down for an arbitrary set of 3 fields.
For the fields appearing in (\ref{eq:expansion}), the interactions at
zero momentum are
\begin{eqnarray}
S_3 & = & \kappa g \left( \phi^3
-\frac{5}{3^2 \sqrt{2}}  B_{\mu}^{\; \mu} \phi^2
-\frac{11}{3^2}  \beta \phi^2
\right.
\label{eq:s3}\\
& &\left.
\hspace{0.6in}
+ \frac{2^4}{3^2}  \phi A_\mu A^{\mu} 
-\frac{5 \cdot 2^3 \sqrt{2}}{3^5}  B_{\mu}^{\; \mu} A_\nu A^\nu
+ \frac{2^8 \sqrt{2}}{3^5}  B^{\mu \nu} A_\mu A_\nu
-\frac{11 \cdot 2^4}{3^5}  \beta A_\mu A^\mu + \cdots \right)
\nonumber
\end{eqnarray}
where
\[
\kappa = \frac{3^{7/2}}{2^7} 
\]

\subsection{Terms quartic in tachyon field}
\label{sec:tachyon}

\begin{figure}
\centering
\begin{picture}(200,100)(- 100,- 50)
\put(-40,0){\line(1,0){80}}
\put(-40,0){\line( -1, -1){30}}
\put(-40,0){\line( -1, 1){30}}
\put(40,0){\line(1, 1){30}}
\put(40,0){\line(1, -1){30}}
\put(0, 7){\makebox(0,0){$\psi$}}
\put(80,30){\makebox(0,0){$\phi$}}
\put(80,-30){\makebox(0,0){$\phi$}}
\put(-80,30){\makebox(0,0){$\phi$}}
\put(-80,-30){\makebox(0,0){$\phi$}}
\end{picture}
\caption[x]{\footnotesize  Tree diagram contributing to $\phi^4$ term}
\label{f:h-diagram}
\end{figure}
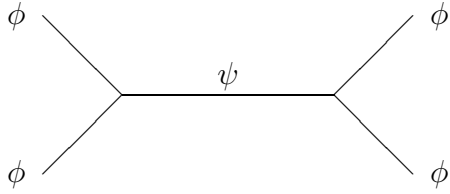

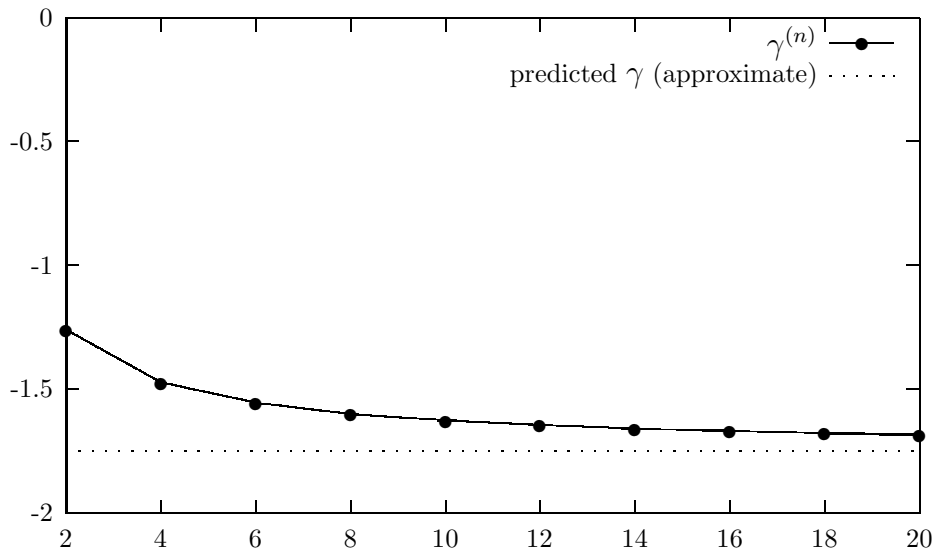
\begin{figure}[htp]
\begin{center}
\setlength{\unitlength}{0.240900pt}
\ifx\plotpoint\undefined\newsavebox{\plotpoint}\fi
\sbox{\plotpoint}{\rule[-0.200pt]{0.400pt}{0.400pt}}%
\begin{picture}(1500,900)(0,0)
\font\gnuplot=cmr10 at 10pt
\gnuplot
\sbox{\plotpoint}{\rule[-0.200pt]{0.400pt}{0.400pt}}%
\put(120.0,82.0){\rule[-0.200pt]{4.818pt}{0.400pt}}
\put(100,82){\makebox(0,0)[r]{-2}}
\put(1440.0,82.0){\rule[-0.200pt]{4.818pt}{0.400pt}}
\put(120.0,277.0){\rule[-0.200pt]{4.818pt}{0.400pt}}
\put(100,277){\makebox(0,0)[r]{-1.5}}
\put(1440.0,277.0){\rule[-0.200pt]{4.818pt}{0.400pt}}
\put(120.0,471.0){\rule[-0.200pt]{4.818pt}{0.400pt}}
\put(100,471){\makebox(0,0)[r]{-1}}
\put(1440.0,471.0){\rule[-0.200pt]{4.818pt}{0.400pt}}
\put(120.0,666.0){\rule[-0.200pt]{4.818pt}{0.400pt}}
\put(100,666){\makebox(0,0)[r]{-0.5}}
\put(1440.0,666.0){\rule[-0.200pt]{4.818pt}{0.400pt}}
\put(120.0,860.0){\rule[-0.200pt]{4.818pt}{0.400pt}}
\put(100,860){\makebox(0,0)[r]{0}}
\put(1440.0,860.0){\rule[-0.200pt]{4.818pt}{0.400pt}}
\put(120.0,82.0){\rule[-0.200pt]{0.400pt}{4.818pt}}
\put(120,41){\makebox(0,0){2}}
\put(120.0,840.0){\rule[-0.200pt]{0.400pt}{4.818pt}}
\put(269.0,82.0){\rule[-0.200pt]{0.400pt}{4.818pt}}
\put(269,41){\makebox(0,0){4}}
\put(269.0,840.0){\rule[-0.200pt]{0.400pt}{4.818pt}}
\put(418.0,82.0){\rule[-0.200pt]{0.400pt}{4.818pt}}
\put(418,41){\makebox(0,0){6}}
\put(418.0,840.0){\rule[-0.200pt]{0.400pt}{4.818pt}}
\put(567.0,82.0){\rule[-0.200pt]{0.400pt}{4.818pt}}
\put(567,41){\makebox(0,0){8}}
\put(567.0,840.0){\rule[-0.200pt]{0.400pt}{4.818pt}}
\put(716.0,82.0){\rule[-0.200pt]{0.400pt}{4.818pt}}
\put(716,41){\makebox(0,0){10}}
\put(716.0,840.0){\rule[-0.200pt]{0.400pt}{4.818pt}}
\put(864.0,82.0){\rule[-0.200pt]{0.400pt}{4.818pt}}
\put(864,41){\makebox(0,0){12}}
\put(864.0,840.0){\rule[-0.200pt]{0.400pt}{4.818pt}}
\put(1013.0,82.0){\rule[-0.200pt]{0.400pt}{4.818pt}}
\put(1013,41){\makebox(0,0){14}}
\put(1013.0,840.0){\rule[-0.200pt]{0.400pt}{4.818pt}}
\put(1162.0,82.0){\rule[-0.200pt]{0.400pt}{4.818pt}}
\put(1162,41){\makebox(0,0){16}}
\put(1162.0,840.0){\rule[-0.200pt]{0.400pt}{4.818pt}}
\put(1311.0,82.0){\rule[-0.200pt]{0.400pt}{4.818pt}}
\put(1311,41){\makebox(0,0){18}}
\put(1311.0,840.0){\rule[-0.200pt]{0.400pt}{4.818pt}}
\put(1460.0,82.0){\rule[-0.200pt]{0.400pt}{4.818pt}}
\put(1460,41){\makebox(0,0){20}}
\put(1460.0,840.0){\rule[-0.200pt]{0.400pt}{4.818pt}}
\put(120.0,82.0){\rule[-0.200pt]{322.806pt}{0.400pt}}
\put(1460.0,82.0){\rule[-0.200pt]{0.400pt}{187.420pt}}
\put(120.0,860.0){\rule[-0.200pt]{322.806pt}{0.400pt}}
\put(120.0,82.0){\rule[-0.200pt]{0.400pt}{187.420pt}}
\put(1300,820){\makebox(0,0)[r]{$\gamma^{(n)}$}}
\put(1320.0,820.0){\rule[-0.200pt]{24.090pt}{0.400pt}}
\put(120,370){\usebox{\plotpoint}}
\multiput(120.00,368.92)(0.899,-0.499){163}{\rule{0.818pt}{0.120pt}}
\multiput(120.00,369.17)(147.302,-83.000){2}{\rule{0.409pt}{0.400pt}}
\multiput(269.00,285.92)(2.348,-0.497){61}{\rule{1.962pt}{0.120pt}}
\multiput(269.00,286.17)(144.927,-32.000){2}{\rule{0.981pt}{0.400pt}}
\multiput(418.00,253.92)(4.214,-0.495){33}{\rule{3.411pt}{0.119pt}}
\multiput(418.00,254.17)(141.920,-18.000){2}{\rule{1.706pt}{0.400pt}}
\multiput(567.00,235.92)(7.740,-0.491){17}{\rule{6.060pt}{0.118pt}}
\multiput(567.00,236.17)(136.422,-10.000){2}{\rule{3.030pt}{0.400pt}}
\multiput(716.00,225.93)(11.248,-0.485){11}{\rule{8.557pt}{0.117pt}}
\multiput(716.00,226.17)(130.239,-7.000){2}{\rule{4.279pt}{0.400pt}}
\multiput(864.00,218.93)(13.419,-0.482){9}{\rule{10.033pt}{0.116pt}}
\multiput(864.00,219.17)(128.175,-6.000){2}{\rule{5.017pt}{0.400pt}}
\multiput(1013.00,212.95)(33.058,-0.447){3}{\rule{19.967pt}{0.108pt}}
\multiput(1013.00,213.17)(107.558,-3.000){2}{\rule{9.983pt}{0.400pt}}
\multiput(1162.00,209.94)(21.683,-0.468){5}{\rule{15.000pt}{0.113pt}}
\multiput(1162.00,210.17)(117.867,-4.000){2}{\rule{7.500pt}{0.400pt}}
\put(1311,205.17){\rule{29.900pt}{0.400pt}}
\multiput(1311.00,206.17)(86.941,-2.000){2}{\rule{14.950pt}{0.400pt}}
\put(120,370){\raisebox{-.8pt}{\makebox(0,0){$\bullet$}}}
\put(269,287){\raisebox{-.8pt}{\makebox(0,0){$\bullet$}}}
\put(418,255){\raisebox{-.8pt}{\makebox(0,0){$\bullet$}}}
\put(567,237){\raisebox{-.8pt}{\makebox(0,0){$\bullet$}}}
\put(716,227){\raisebox{-.8pt}{\makebox(0,0){$\bullet$}}}
\put(864,220){\raisebox{-.8pt}{\makebox(0,0){$\bullet$}}}
\put(1013,214){\raisebox{-.8pt}{\makebox(0,0){$\bullet$}}}
\put(1162,211){\raisebox{-.8pt}{\makebox(0,0){$\bullet$}}}
\put(1311,207){\raisebox{-.8pt}{\makebox(0,0){$\bullet$}}}
\put(1460,205){\raisebox{-.8pt}{\makebox(0,0){$\bullet$}}}
\put(1370,820){\raisebox{-.8pt}{\makebox(0,0){$\bullet$}}}
\put(1300,769){\makebox(0,0)[r]{predicted $\gamma$ (approximate)}}
\multiput(1320,769)(20.756,0.000){5}{\usebox{\plotpoint}}
\put(1420,769){\usebox{\plotpoint}}
\put(120,179){\usebox{\plotpoint}}
\put(120.00,179.00){\usebox{\plotpoint}}
\put(140.76,179.00){\usebox{\plotpoint}}
\put(161.51,179.00){\usebox{\plotpoint}}
\put(182.27,179.00){\usebox{\plotpoint}}
\put(203.02,179.00){\usebox{\plotpoint}}
\put(223.78,179.00){\usebox{\plotpoint}}
\put(244.53,179.00){\usebox{\plotpoint}}
\put(265.29,179.00){\usebox{\plotpoint}}
\put(286.04,179.00){\usebox{\plotpoint}}
\put(306.80,179.00){\usebox{\plotpoint}}
\put(327.55,179.00){\usebox{\plotpoint}}
\put(348.31,179.00){\usebox{\plotpoint}}
\put(369.07,179.00){\usebox{\plotpoint}}
\put(389.82,179.00){\usebox{\plotpoint}}
\put(410.58,179.00){\usebox{\plotpoint}}
\put(431.33,179.00){\usebox{\plotpoint}}
\put(452.09,179.00){\usebox{\plotpoint}}
\put(472.84,179.00){\usebox{\plotpoint}}
\put(493.60,179.00){\usebox{\plotpoint}}
\put(514.35,179.00){\usebox{\plotpoint}}
\put(535.11,179.00){\usebox{\plotpoint}}
\put(555.87,179.00){\usebox{\plotpoint}}
\put(576.62,179.00){\usebox{\plotpoint}}
\put(597.38,179.00){\usebox{\plotpoint}}
\put(618.13,179.00){\usebox{\plotpoint}}
\put(638.89,179.00){\usebox{\plotpoint}}
\put(659.64,179.00){\usebox{\plotpoint}}
\put(680.40,179.00){\usebox{\plotpoint}}
\put(701.15,179.00){\usebox{\plotpoint}}
\put(721.91,179.00){\usebox{\plotpoint}}
\put(742.66,179.00){\usebox{\plotpoint}}
\put(763.42,179.00){\usebox{\plotpoint}}
\put(784.18,179.00){\usebox{\plotpoint}}
\put(804.93,179.00){\usebox{\plotpoint}}
\put(825.69,179.00){\usebox{\plotpoint}}
\put(846.44,179.00){\usebox{\plotpoint}}
\put(867.20,179.00){\usebox{\plotpoint}}
\put(887.95,179.00){\usebox{\plotpoint}}
\put(908.71,179.00){\usebox{\plotpoint}}
\put(929.46,179.00){\usebox{\plotpoint}}
\put(950.22,179.00){\usebox{\plotpoint}}
\put(970.98,179.00){\usebox{\plotpoint}}
\put(991.73,179.00){\usebox{\plotpoint}}
\put(1012.49,179.00){\usebox{\plotpoint}}
\put(1033.24,179.00){\usebox{\plotpoint}}
\put(1054.00,179.00){\usebox{\plotpoint}}
\put(1074.75,179.00){\usebox{\plotpoint}}
\put(1095.51,179.00){\usebox{\plotpoint}}
\put(1116.26,179.00){\usebox{\plotpoint}}
\put(1137.02,179.00){\usebox{\plotpoint}}
\put(1157.77,179.00){\usebox{\plotpoint}}
\put(1178.53,179.00){\usebox{\plotpoint}}
\put(1199.29,179.00){\usebox{\plotpoint}}
\put(1220.04,179.00){\usebox{\plotpoint}}
\put(1240.80,179.00){\usebox{\plotpoint}}
\put(1261.55,179.00){\usebox{\plotpoint}}
\put(1282.31,179.00){\usebox{\plotpoint}}
\put(1303.06,179.00){\usebox{\plotpoint}}
\put(1323.82,179.00){\usebox{\plotpoint}}
\put(1344.57,179.00){\usebox{\plotpoint}}
\put(1365.33,179.00){\usebox{\plotpoint}}
\put(1386.09,179.00){\usebox{\plotpoint}}
\put(1406.84,179.00){\usebox{\plotpoint}}
\put(1427.60,179.00){\usebox{\plotpoint}}
\put(1448.35,179.00){\usebox{\plotpoint}}
\put(1460,179){\usebox{\plotpoint}}
\end{picture}
\end{center}
\caption[x]{\footnotesize Approximations to coefficient of $\phi^4$ term}
\label{f:tachyon}
\end{figure}

Kostelecky and Samuel proposed in \cite{ks-open} that the answers to
many physical questions in string field theory might be very well
approximated by truncating the string field action at a finite level.
As a first check of this proposal they considered the effective
quartic term in the tachyon field $\phi^4$ which arises from
integrating out all the massive scalar fields in the theory.  For each
massive scalar field $\psi$ whose quadratic term and coupling to $\phi^2$
are given by
\begin{equation}
S_\psi =  \frac{a}{2}  \psi^2 + c \psi \phi^2
\end{equation}
there is a term in the effective potential for $\phi$ of the form
\[
S_{\phi 4} = -\frac{c^2}{2a}  \phi^4.
\]
The (tree-level) diagram for such a term is shown in
Figure~\ref{f:h-diagram}.
Thus, for example, the contributions from the level 2 fields $\beta,
B_{\mu}^{\; \mu}$ to the $\phi^4$ term in the effective tachyon potential
are seen from (\ref{eq:s2},\ref{eq:s3}) to be
\begin{equation}
\kappa^2 g^2 \alpha' \left(\frac{1}{2} (\frac{11}{3^2})^2
-26 \cdot \frac{1}{2}  (\frac{5}{3^2 \sqrt{2}})^2 \right) \phi^4
=-\kappa^2 g^2 \alpha' \left(\frac{34}{27}  \right) \phi^4.
\label{eq:t4-2}
\end{equation}
The exact quartic term in the tachyon potential was determined in
\cite{ks-exact}, and is given by
\[
\kappa^2 g^2  \alpha' \gamma \phi^4
\]
where
\begin{equation}
\gamma \approx -1.75 \pm 0.02
\label{eq:t4-total}
\end{equation}
The quartic term produced by integrating out the level 2 fields
(\ref{eq:t4-2}) represents 72\% of the total (\ref{eq:t4-total}).  In
\cite{ks-open}, Kostelecky and Samuel calculated the contribution from
level 4 fields, and showed that these produce an additional 12\% of
the total quartic term.  As a first test of how well the level
truncation method does in reproducing quartic terms in the effective
action on a D-brane, and as a test of our algorithm for calculating
contributions from fields at arbitrary level, we have used {\it
Mathematica} to carry out this truncation exactly up to order 20.  The
results of this calculation are summarized in Table~\ref{t:t4}.  
\begin{table}[htp]
\begin{center}
\begin{tabular}{| r | r | c | r | c |}
\hline
\hline
$n$& \# of fields & $\hat{\gamma}^{(n)}$ &
$\gamma^{(n)}$ & $\gamma^{(n)}/\gamma_{{\rm exact}}$\\
\hline
\hline
2 & 2 & 
 \rule[-0.1cm]{0cm}{0.56cm} 
$-\frac{
2^{}
\cdot
17^{}
}{
3^{3}
}$
 $\approx
 -1.25926$ & $ -1.259$
&   0.72 $\pm 0.01$\\
4 & 7 & 
 \rule[-0.1cm]{0cm}{0.56cm} 
$-\frac{
1399^{}
}{
3^{8}
}$
 $\approx
 -0.21323$ & $ -1.472$
&   0.84 $\pm 0.01$\\
6 & 20 & 
 \rule[-0.1cm]{0cm}{0.56cm} 
$-\frac{
2^{2}
\cdot
7^{}
\cdot
643463^{}
}{
3^{16}
\cdot
5^{}
}$
 $\approx
 -0.08371$ & $ -1.556$
&   0.89 $\pm 0.01$\\
8 & 55 & 
 \rule[-0.1cm]{0cm}{0.56cm} 
$-\frac{
167^{}
\cdot
1846847^{}
}{
2^{}
\cdot
3^{20}
}$
 $\approx
 -0.04423$ & $ -1.600$
&   0.92 $\pm 0.01$\\
10 & 139 & 
 \rule[-0.1cm]{0cm}{0.56cm} 
$-\frac{
2^{}
\cdot
5^{}
\cdot
193^{}
\cdot
241^{}
\cdot
1341187^{}
}{
3^{28}
}$
 $\approx
 -0.02727$ & $ -1.628$
&   0.93 $\pm 0.01$\\
12 & 331 & 
 \rule[-0.1cm]{0cm}{0.56cm} 
$-\frac{
5^{}
\cdot
2912243^{}
\cdot
232794533^{}
}{
3^{34}
\cdot
11^{}
}$
 $\approx
 -0.01848$ & $ -1.646$
&   0.94 $\pm 0.01$\\
14 & 747 & 
 \rule[-0.1cm]{0cm}{0.56cm} 
$-\frac{
2^{2}
\cdot
11^{}
\cdot
1775214529932629^{}
}{
3^{37}
\cdot
13^{}
}$
 $\approx
 -0.01334$ & $ -1.660$
&   0.95 $\pm 0.01$\\
16 & 1618 & 
 \rule[-0.1cm]{0cm}{0.56cm} 
$-\frac{
298001292970739836603^{}
}{
2^{}
\cdot
3^{45}
\cdot
5^{}
}$
 $\approx
 -0.01009$ & $ -1.670$
&   0.96 $\pm 0.01$\\
18 & 3375 & 
 \rule[-0.1cm]{0cm}{0.56cm} 
$-\frac{
2^{2}
\cdot
7^{}
\cdot
21159416989^{}
\cdot
1463230224529^{}
}{
3^{52}
\cdot
17^{}
}$
 $\approx
 -0.00789$ & $ -1.677$
&   0.96 $\pm 0.01$\\
20 & 6818 & 
 \rule[-0.1cm]{0cm}{0.56cm} 
$-\frac{
31^{}
\cdot
1061^{}
\cdot
1319^{}
\cdot
164809^{}
\cdot
26468211709651^{}
}{
3^{57}
\cdot
19^{}
}$
 $\approx
 -0.00634$ & $ -1.684$
&   0.96 $\pm 0.01$\\
\hline
\hline
\end{tabular}
\caption[x]{\footnotesize Contributions at each level to coefficient
of $\phi^4$}
\label{t:t4}
\end{center}
\end{table}
A more
detailed description of the contributions from each field up to level
6 is given in Appendix B.  We denote by $\hat{\gamma}^{(n)}$ the
contribution to $\gamma$ from fields at level $n$, and by
$\gamma^{(n)}$ the cumulative $\gamma$ found by adding the
contribution from all fields at levels $m \leq n$.  After summing the
contributions from all 13,112 fields at levels $\leq 20$, we find that
the exact result from \cite{ks-exact} is reproduced to within approximately
$4\%$.
A graph of
the coefficient of the quartic term coming from the successive level
truncation approximations is shown in Figure~\ref{f:tachyon}.
There are several significant aspects of how the successive
approximations to this term in the effective action behave as the
level number increases.  First, the approximations seem to be
completely convergent, with no sign of bad behavior at high level as
one might expect if the series were only asymptotic but not
convergent.  Second, the successive approximations seem to have
slower than exponential falloff, but do not seem to follow a simple
power law.  It would be very interesting to have a better
understanding of the convergence properties of these terms from a
study of the asymptotic properties of the states and their couplings.

\subsection{Terms quartic in gauge field}
\label{sec:gauge}

Now let us turn to the massless gauge field $A^\mu$ associated with
the spin 1 states $\alpha^\mu_{-1}| 0 \rangle$.  Again, we are
interested in finding quartic terms in the gauge field.  Let us first
consider the abelian U(1) theory associated with a single bosonic
D-brane.  In this theory the only Lorentz invariant term which is
quartic in $A^\mu$ is $(A^\mu A_\mu)^2$.  Such a term is indeed
induced by integrating out the tachyon at tree level.  From the
vertices in (\ref{eq:s3}) of the form $\phi A^\mu A_\mu$ we see that
there is a term in the effective action of the U(1) theory
\begin{equation}
\kappa^2 g^2 \alpha' \left(\frac{1}{2} (\frac{2^4}{3^2})^2
\right)   (A^\mu A_\mu)^2
=\kappa^2 g^2 \alpha' \left(\frac{128}{81} \right)
  (A^\mu A_\mu)^2
\label{eq:a4-0}
\end{equation}
The appearance of
this term in the effective action for the world-volume gauge field may
seem rather surprising, as this term does not appear as the 0-momentum
part of any expression which is invariant under the U(1) gauge
symmetry we expect to have in the effective gauge theory found after
integrating out the tachyonic and massive string modes.  Indeed, the
gauge noninvariance of this term suggests that contributions to the
quartic term from higher string modes should cancel this term, so that
in the true effective U(1) gauge theory there is no $A^4$ term.
This argument is not completely conclusive, however, since by choosing
Feynman-Siegel gauge we have broken the gauge invariance of the
effective U(1) theory.  For example, the kinetic term for the vector
field in Feynman-Siegel gauge is simply $\frac{1}{2} \partial_\mu
A_\nu \partial^\mu A^\nu $, which is the kinetic term for the massless
vector field in Lorentz gauge $\partial_\mu A^\mu = 0$.

A related argument for the conclusion that the $A^4$ term should
vanish in the abelian theory is that if we compactify the theory in
some set of dimensions and then perform a series of T-duality
symmetries which change the Neumann boundary conditions on the open
string to Dirichlet boundary conditions in those directions, we have
an effective theory of a bosonic D-brane of dimension $p < 25$
whose position in the transverse spatial directions is
described by a set of  coordinates $X^\mu$ which are T-dual to the
constant gauge field components $A^\mu$ giving rise to nontrivial
holonomies around the compact directions in
original picture.  The  quartic term in the T-dual picture analogous to
(\ref{eq:a4-0}) is of the form $(X^\mu X_\mu)^2$; this term breaks
translation invariance, perhaps even a worse transgression than the
breaking of gauge invariance by (\ref{eq:a4-0}).

It may seem strange that we have integrated out the tachyon field and
kept the massless vector field.  The usual approach to deriving an
effective action in a theory with fields at many energy scales is to
integrate out the fields with very large masses keeping the light or
massless fields.  This gives a low-energy effective theory for the
light fields in the theory which has a clear physical significance.
It is less clear how to interpret the action for the vector field on a
bosonic D-brane which arises when we integrate out the tachyon, which
has negative mass squared.  Because the energy scale of the tachyon is
the same as that of the massive string states, however, it seems most
reasonable to integrate out the tachyon field along with the massive
states when constructing an effective action for the vector field.  As
we will discuss in the next section, we expect that this effective
action should take the form of a bosonic Born-Infeld action.

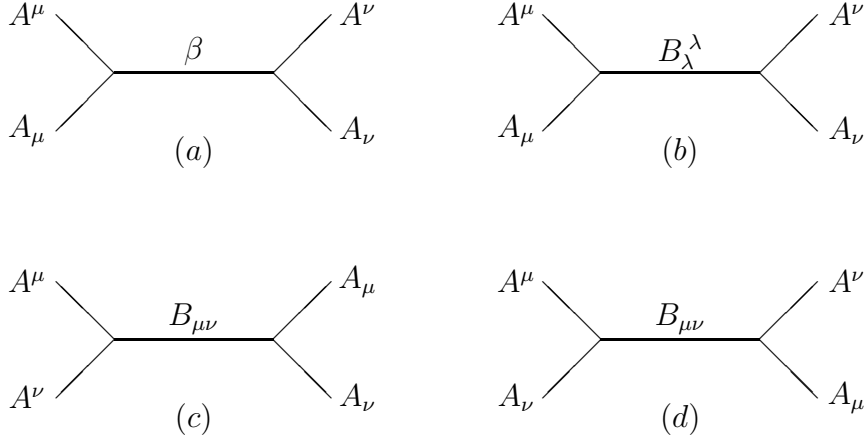
\begin{figure}
\centering
\begin{picture}(320,100)(- 160,- 100)
\put(-122,-62){\line(1, 0){60}}
\put(-122,-62){\line( -1, -1){22}}
\put(-122,-62){\line( -1, 1){22}}
\put(-62,-62){\line(1, 1){22}}
\put(-62,-62){\line(1, -1){22}}
\put(-92, -54){\makebox(0,0){$\beta$}}
\put(-155,-40){\makebox(0,0){$A^\mu$}}
\put(-155,-84){\makebox(0,0){$A_\mu$}}
\put(-30,-40){\makebox(0,0){$A^\nu$}}
\put(-30,-84){\makebox(0,0){$A_\nu$}}
\put(-92, -92,){\makebox(0,0){$(a)$}}

\put(122,-62){\line(-1, 0){60}}
\put(122,-62){\line(1, -1){22}}
\put(122,-62){\line(1, 1){22}}
\put(62,-62){\line(-1, 1){22}}
\put(62,-62){\line(-1, -1){22}}
\put(92, -54){\makebox(0,0){$B_\lambda^{\; \lambda}$}}
\put(155,-40){\makebox(0,0){$A^\nu$}}
\put(155,-84){\makebox(0,0){$A_\nu$}}
\put(30,-40){\makebox(0,0){$A^\mu$}}
\put(30,-84){\makebox(0,0){$A_\mu$}}
\put(92, -92,){\makebox(0,0){$(b)$}}
\end{picture}
\begin{picture}(320,100)(- 160,- 100)
\put(-122,-62){\line(1, 0){60}}
\put(-122,-62){\line( -1, -1){22}}
\put(-122,-62){\line( -1, 1){22}}
\put(-62,-62){\line(1, 1){22}}
\put(-62,-62){\line(1, -1){22}}
\put(-92, -54){\makebox(0,0){$B_{\mu \nu}$}}
\put(-155,-40){\makebox(0,0){$A^\mu$}}
\put(-155,-84){\makebox(0,0){$A^\nu$}}
\put(-30,-40){\makebox(0,0){$A_\mu$}}
\put(-30,-84){\makebox(0,0){$A_\nu$}}
\put(-92, -92,){\makebox(0,0){$(c)$}}

\put(122,-62){\line(-1, 0){60}}
\put(122,-62){\line(1, -1){22}}
\put(122,-62){\line(1, 1){22}}
\put(62,-62){\line(-1, 1){22}}
\put(62,-62){\line(-1, -1){22}}
\put(92, -54){\makebox(0,0){$B_{\mu \nu}$}}
\put(155,-40){\makebox(0,0){$A^\nu$}}
\put(155,-84){\makebox(0,0){$A_\mu$}}
\put(30,-40){\makebox(0,0){$A^\mu$}}
\put(30,-84){\makebox(0,0){$A_\nu$}}
\put(92, -92,){\makebox(0,0){$(d)$}}
\end{picture}
\caption[x]{\footnotesize Diagrams contributing to terms quartic in
vector field at level 2}
\label{f:t4-b}
\end{figure}

In any case, the effective action which appears when integrating out
the tachyon as well as the massive fields in the theory seems to be
quite well behaved, at least at quartic order.  A consideration of the
effects of the fields of level 2 and higher bears out the conjecture
that the term (\ref{eq:a4-0}) is completely cancelled by the infinite
tower of massive string states.  At level 2, there are contributions
to the quartic gauge field term from the fields $\beta, B_{\mu \nu}$.
The diagrams associated with these contributions are depicted in
Figure~\ref{f:t4-b}. 
The total contribution from these diagrams to
the quartic term in the gauge field is
\begin{equation}
-\kappa^2 g^2 \alpha' \left(\frac{71168}{59049}  \right) (A^\mu A_\mu)^2.
\label{eq:a4-2}
\end{equation}
This term cancels $76\%$ of the spurious term generated by the
tachyon.  It is instructive to consider how this term is computed,
so we include some of the details in this particular case.  Let us
consider in particular those terms with two distinct pairs of
indices on the gauge field, such as $A_i A_i A_j A_j$ with $i \neq j$.
For terms of this type, the field $\beta$ and the diagonal components
$B_{\mu }^{\; \mu}$ with $\mu \neq i, j$
contribute in the same fashion as we have found
previously through the couplings of the forms $\beta (A^\mu A_\mu)$
and $B_{\nu}^{\; \nu} (A^\mu A_\mu)$ (diagrams (a) and (b) in
Figure~\ref{f:t4-b}), giving terms in the effective action
\begin{equation}
\kappa^2 g^2 \alpha' \left(\frac{1}{2} (\frac{11 \cdot 2^4}{3^5})^2
-24 \cdot \frac{1}{2}  (\frac{5 \cdot 2^{7/2}}{3^5})^2 \right) 
\left( 2 A_i^2 A_j^2\right)
=-\kappa^2 g^2 \alpha' \left(\frac{22912}{59049}  \right) 
\left( 2 A_i A_i A_j A_j\right).
\end{equation}
The fields $B_{ii}$ and $B_{jj}$ 
together contribute
\begin{equation}
\kappa^2 g^2 \alpha' \left(
-2 \cdot  (\frac{(32-5) \cdot 2^{7/2}}{3^5})
(-\frac{5 \cdot 2^{7/2}}{3^5}) \right) 
\left(A_i^2 A_j^2\right)
=\kappa^2 g^2 \alpha' \left(\frac{17280}{59049}  \right) 
\left( 2 A_i A_i A_j A_j\right).
\end{equation}
In addition to these terms, there are terms arising from the field
$B_{ij}$, associated with diagrams (c)
and (d) in Figure ~\ref{f:t4-b}.  These additional terms come to
\begin{equation}
\kappa^2 g^2 \alpha' \left(
- \frac{1}{2}  (\frac{2 \cdot 2^8\sqrt{2}}{3^5})^2 \right) 
\left(A_i A_j\right)^2
=-\kappa^2 g^2 \alpha' \left(\frac{65536}{59049}  \right) 
\left( 2 A_i A_i A_j A_j\right).
\end{equation}
Summing these, we arrive at the total coefficient described by
(\ref{eq:a4-2}).   We have performed a similar analysis for the terms of
the form $A_i^4$, and find the same result for the overall
coefficient, as is necessitated by Lorentz invariance.

We have thus explicitly shown that by including the fields at level 2,
we cancel 76\% of the spurious $A^4$ term in the U(1) theory induced
by the tachyon.  We have continued this calculation further using {\it
Mathematica}, determining the exact contribution to the quartic term
arising from all fields up to level 16.  Indeed, we find that the
contributions at each level serve to decrease the total coefficient of
the quartic term, so that the term arising from the tachyon has been
97\% cancelled at level 16.  Before presenting the results of this
calculation, we briefly discuss the modifications which arise
in the nonabelian theory.

Let us consider the nonabelian theory which arises when we add
Chan-Paton factors to the strings in the original string field theory.
If the Chan-Paton factors range from 1 through $N$ then all the fields
in the string field expansion (\ref{eq:expansion}) become $N \times N$
matrices.  The gauge field, in particular, becomes a U(N) gauge field.
We expect that the effective action for this gauge field will have a
leading term of the form $F^2$, which in the 0-momentum theory reduces
to a term of the form
\begin{equation}
[A_\mu, A_\nu]^2
\label{eq:f2}
\end{equation}
Thus, we should hope that by keeping track of the nonabelian structure
of the fields in the U(N) theory, we should find that instead of
vanishing identically, the quartic term in the effective gauge field
action will take the form (\ref{eq:f2}).  If string field theory is
sufficiently well behaved, we might hope to find that summing the
contributions to this term from successive level truncation will give
us a series of terms which converge to a finite value for the
coefficient of the term (\ref{eq:f2}).

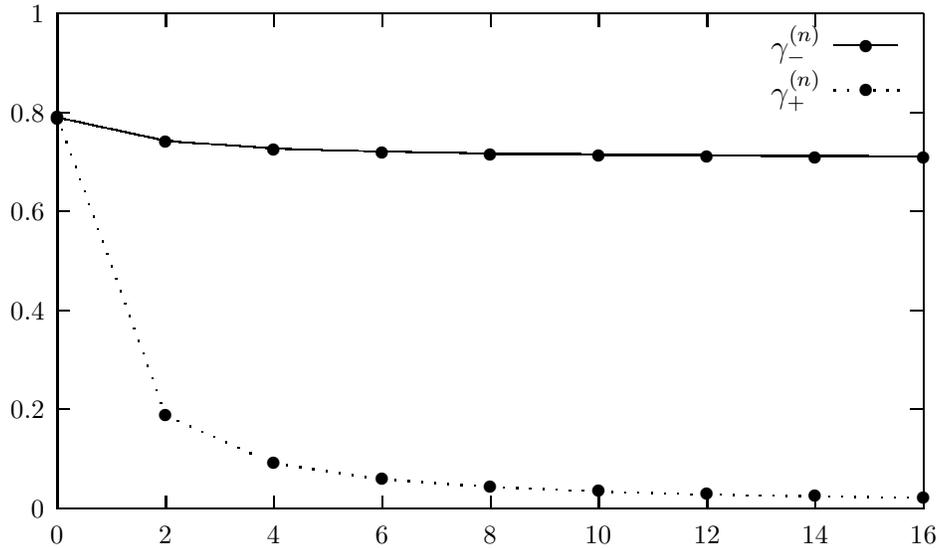
\begin{figure}[htp]
\begin{center}
\setlength{\unitlength}{0.240900pt}
\ifx\plotpoint\undefined\newsavebox{\plotpoint}\fi
\sbox{\plotpoint}{\rule[-0.200pt]{0.400pt}{0.400pt}}%
\begin{picture}(1500,900)(0,0)
\font\gnuplot=cmr10 at 10pt
\gnuplot
\sbox{\plotpoint}{\rule[-0.200pt]{0.400pt}{0.400pt}}%
\put(100.0,82.0){\rule[-0.200pt]{4.818pt}{0.400pt}}
\put(80,82){\makebox(0,0)[r]{0}}
\put(1440.0,82.0){\rule[-0.200pt]{4.818pt}{0.400pt}}
\put(100.0,238.0){\rule[-0.200pt]{4.818pt}{0.400pt}}
\put(80,238){\makebox(0,0)[r]{0.2}}
\put(1440.0,238.0){\rule[-0.200pt]{4.818pt}{0.400pt}}
\put(100.0,393.0){\rule[-0.200pt]{4.818pt}{0.400pt}}
\put(80,393){\makebox(0,0)[r]{0.4}}
\put(1440.0,393.0){\rule[-0.200pt]{4.818pt}{0.400pt}}
\put(100.0,549.0){\rule[-0.200pt]{4.818pt}{0.400pt}}
\put(80,549){\makebox(0,0)[r]{0.6}}
\put(1440.0,549.0){\rule[-0.200pt]{4.818pt}{0.400pt}}
\put(100.0,704.0){\rule[-0.200pt]{4.818pt}{0.400pt}}
\put(80,704){\makebox(0,0)[r]{0.8}}
\put(1440.0,704.0){\rule[-0.200pt]{4.818pt}{0.400pt}}
\put(100.0,860.0){\rule[-0.200pt]{4.818pt}{0.400pt}}
\put(80,860){\makebox(0,0)[r]{1}}
\put(1440.0,860.0){\rule[-0.200pt]{4.818pt}{0.400pt}}
\put(100.0,82.0){\rule[-0.200pt]{0.400pt}{4.818pt}}
\put(100,41){\makebox(0,0){0}}
\put(100.0,840.0){\rule[-0.200pt]{0.400pt}{4.818pt}}
\put(270.0,82.0){\rule[-0.200pt]{0.400pt}{4.818pt}}
\put(270,41){\makebox(0,0){2}}
\put(270.0,840.0){\rule[-0.200pt]{0.400pt}{4.818pt}}
\put(440.0,82.0){\rule[-0.200pt]{0.400pt}{4.818pt}}
\put(440,41){\makebox(0,0){4}}
\put(440.0,840.0){\rule[-0.200pt]{0.400pt}{4.818pt}}
\put(610.0,82.0){\rule[-0.200pt]{0.400pt}{4.818pt}}
\put(610,41){\makebox(0,0){6}}
\put(610.0,840.0){\rule[-0.200pt]{0.400pt}{4.818pt}}
\put(780.0,82.0){\rule[-0.200pt]{0.400pt}{4.818pt}}
\put(780,41){\makebox(0,0){8}}
\put(780.0,840.0){\rule[-0.200pt]{0.400pt}{4.818pt}}
\put(950.0,82.0){\rule[-0.200pt]{0.400pt}{4.818pt}}
\put(950,41){\makebox(0,0){10}}
\put(950.0,840.0){\rule[-0.200pt]{0.400pt}{4.818pt}}
\put(1120.0,82.0){\rule[-0.200pt]{0.400pt}{4.818pt}}
\put(1120,41){\makebox(0,0){12}}
\put(1120.0,840.0){\rule[-0.200pt]{0.400pt}{4.818pt}}
\put(1290.0,82.0){\rule[-0.200pt]{0.400pt}{4.818pt}}
\put(1290,41){\makebox(0,0){14}}
\put(1290.0,840.0){\rule[-0.200pt]{0.400pt}{4.818pt}}
\put(1460.0,82.0){\rule[-0.200pt]{0.400pt}{4.818pt}}
\put(1460,41){\makebox(0,0){16}}
\put(1460.0,840.0){\rule[-0.200pt]{0.400pt}{4.818pt}}
\put(100.0,82.0){\rule[-0.200pt]{327.624pt}{0.400pt}}
\put(1460.0,82.0){\rule[-0.200pt]{0.400pt}{187.420pt}}
\put(100.0,860.0){\rule[-0.200pt]{327.624pt}{0.400pt}}
\put(100.0,82.0){\rule[-0.200pt]{0.400pt}{187.420pt}}
\put(1300,810){\makebox(0,0)[r]{\shortstack{$\gamma_-^{(n)}$}}}
\put(1320.0,810.0){\rule[-0.200pt]{24.090pt}{0.400pt}}
\put(100,697){\usebox{\plotpoint}}
\multiput(100.00,695.92)(2.314,-0.498){71}{\rule{1.938pt}{0.120pt}}
\multiput(100.00,696.17)(165.978,-37.000){2}{\rule{0.969pt}{0.400pt}}
\multiput(270.00,658.92)(7.303,-0.492){21}{\rule{5.767pt}{0.119pt}}
\multiput(270.00,659.17)(158.031,-12.000){2}{\rule{2.883pt}{0.400pt}}
\multiput(440.00,646.93)(18.855,-0.477){7}{\rule{13.700pt}{0.115pt}}
\multiput(440.00,647.17)(141.565,-5.000){2}{\rule{6.850pt}{0.400pt}}
\multiput(610.00,641.95)(37.747,-0.447){3}{\rule{22.767pt}{0.108pt}}
\multiput(610.00,642.17)(122.747,-3.000){2}{\rule{11.383pt}{0.400pt}}
\put(780,638.17){\rule{34.100pt}{0.400pt}}
\multiput(780.00,639.17)(99.224,-2.000){2}{\rule{17.050pt}{0.400pt}}
\put(950,636.67){\rule{40.953pt}{0.400pt}}
\multiput(950.00,637.17)(85.000,-1.000){2}{\rule{20.476pt}{0.400pt}}
\put(1120,635.67){\rule{40.953pt}{0.400pt}}
\multiput(1120.00,636.17)(85.000,-1.000){2}{\rule{20.476pt}{0.400pt}}
\put(1290,634.67){\rule{40.953pt}{0.400pt}}
\multiput(1290.00,635.17)(85.000,-1.000){2}{\rule{20.476pt}{0.400pt}}
\put(100,697){\raisebox{-.8pt}{\makebox(0,0){$\bullet$}}}
\put(270,660){\raisebox{-.8pt}{\makebox(0,0){$\bullet$}}}
\put(440,648){\raisebox{-.8pt}{\makebox(0,0){$\bullet$}}}
\put(610,643){\raisebox{-.8pt}{\makebox(0,0){$\bullet$}}}
\put(780,640){\raisebox{-.8pt}{\makebox(0,0){$\bullet$}}}
\put(950,638){\raisebox{-.8pt}{\makebox(0,0){$\bullet$}}}
\put(1120,637){\raisebox{-.8pt}{\makebox(0,0){$\bullet$}}}
\put(1290,636){\raisebox{-.8pt}{\makebox(0,0){$\bullet$}}}
\put(1460,635){\raisebox{-.8pt}{\makebox(0,0){$\bullet$}}}
\put(1370,810){\raisebox{-.8pt}{\makebox(0,0){$\bullet$}}} 
\put(1300,739){\makebox(0,0)[r]{\shortstack{$\gamma_+^{(n)}$}}} 
\multiput(1320,739)(20.756,0.000){5}{\usebox{\plotpoint}}
\put(1420,739){\usebox{\plotpoint}}
\put(100,697){\usebox{\plotpoint}}
\multiput(100,697)(7.073,-19.513){25}{\usebox{\plotpoint}}
\multiput(270,228)(18.990,-8.378){8}{\usebox{\plotpoint}}
\multiput(440,153)(20.535,-3.020){9}{\usebox{\plotpoint}}
\multiput(610,128)(20.704,-1.461){8}{\usebox{\plotpoint}}
\multiput(780,116)(20.738,-0.854){8}{\usebox{\plotpoint}}
\multiput(950,109)(20.747,-0.610){8}{\usebox{\plotpoint}}
\multiput(1120,104)(20.752,-0.366){9}{\usebox{\plotpoint}}
\multiput(1290,101)(20.754,-0.244){8}{\usebox{\plotpoint}}
\put(1460,99){\usebox{\plotpoint}}
\put(100,697){\makebox(0,0){$\bullet$}}
\put(270,228){\makebox(0,0){$\bullet$}}
\put(440,153){\makebox(0,0){$\bullet$}}
\put(610,128){\makebox(0,0){$\bullet$}}
\put(780,116){\makebox(0,0){$\bullet$}}
\put(950,109){\makebox(0,0){$\bullet$}}
\put(1120,104){\makebox(0,0){$\bullet$}}
\put(1290,101){\makebox(0,0){$\bullet$}}
\put(1460,99){\makebox(0,0){$\bullet$}}
\put(1370,739){\makebox(0,0){$\bullet$}}
\end{picture}
\end{center}
\caption[x]{\footnotesize Approximations to coefficients of $A^4$ terms}
\label{f:gauge}
\end{figure}

In the nonabelian theory with Chan-Paton indices, the quadratic and
cubic terms (\ref{eq:s2},\ref{eq:s3}) in the string field action are
modified by simply taking the trace of each term.  It is important to
note, however, that this trace must be taken before the order of the
fields in the term is modified.  Because the Neumann coefficients
$N^{rs}_{nm}$ are not necessarily symmetric under a reversal of order
in the 3 strings at a string vertex, it is necessary to separately
compute the coupling for each ordering of the fields in the vertex.
For the terms describing the couplings of the level 2 fields to a pair
of gauge fields, however, this subtlety is not important, and the
relevant coupling terms are simply
\begin{equation}
-\frac{5 \cdot 2^3 \sqrt{2}}{3^5}  
{\rm Tr}\; \left(B_{\mu}^{\; \mu} A_\nu A^\nu \right)
+ \frac{2^7 \sqrt{2}}{3^5}  
{\rm Tr}\; \left(B^{\mu \nu} A_\nu A_\mu +B^{\mu \nu} A_\mu A_\nu\right)
-\frac{11 \cdot 2^4}{3^5}
{\rm Tr}\; \left(  \beta A_\mu A^\mu \right)
\end{equation}
We can now go through the calculation of the terms quartic in $A^\mu$
for the nonabelian theory just as we did above in the abelian case.
The term (\ref{eq:a4-0}) produced by integrating out the tachyon is
unchanged except for the addition of an overall trace.
Now, however, we find that the term (\ref{eq:a4-2}) decomposes into
a sum of two terms with distinct ordering of the $A$'s
\begin{equation}
\kappa^2 g^2 \alpha' 
\left[  \gamma_1 {\rm Tr}\; \left( A^\mu A_\mu A^\nu A_\nu \right)
+ \gamma_2 {\rm Tr}\; \left( A^\mu A^\nu A_\mu A_\nu \right) \right]
\label{eq:quartic-g}
\end{equation}
where
\begin{eqnarray*}
\gamma_1  =  -\frac{38400}{59049}  \;\;\;\;\; & & \;\;\;\;\;
\gamma_2  =  -\frac{32768}{59049} 
\end{eqnarray*}
In this decomposition, the diagrams (a, b, c) of Figure~\ref{f:t4-b}
contribute to $\gamma_1$ while diagram (d) contributes to $\gamma_2$.
In general, this will be the situation at every level, with nonzero
contributions to both coefficients $\gamma_1$ and $\gamma_2$.  We have
computed these coefficients from terms at all levels up to 16.
These results are summarized in Table 2.  
\begin{table}[htp]
\begin{center}
\begin{tabular}{| r | c | c  | c | c| c | c |}
\hline
\hline
$n$ & $\hat{\gamma}_1^{(n)}$ & $\hat{\gamma}_2^{(n)}$ &
$\gamma_1^{(n)}$ & $\gamma_2^{(n)}$ &
$\gamma_+^{(n)}$ &
$\gamma_-^{(n)}$\\
\hline
\hline
table
 \rule[-0.1cm]{0cm}{0.56cm} 
0 & 
$\frac{
2^{7}
}{
3^{4}
}$
$ \approx  1.58020$ & 
0
$ \approx  0.00000$ & $  1.580$ & $  0.000$ & $ 0.7901$ & $ -0.790$\\
 \rule[-0.1cm]{0cm}{0.56cm} 
2 & 
$-\frac{
2^{9}
\cdot
5^{2}
}{
3^{9}
}$
$ \approx -0.65031$ & 
$-\frac{
2^{15}
}{
3^{10}
}$
$ \approx -0.55493$ & $  0.930$ & $ -0.555$ & $ 0.1875$ & $ -0.742$\\
 \rule[-0.1cm]{0cm}{0.56cm} 
4 & 
$-\frac{
2^{8}
\cdot
6229^{}
}{
3^{15}
}$
$ \approx -0.11113$ & 
$-\frac{
2^{17}
}{
3^{13}
}$
$ \approx -0.08221$ & $  0.819$ & $ -0.637$ & $ 0.0908$ & $ -0.728$\\
 \rule[-0.1cm]{0cm}{0.56cm} 
6 & 
$ \approx -0.03878$ & 
$ \approx -0.02528$ & $  0.780$ & $ -0.662$ & $ 0.0588$ & $ -0.721$\\
 \rule[-0.1cm]{0cm}{0.56cm} 
8 & 
$ \approx -0.01933$ & 
$ \approx -0.01150$ & $  0.761$ & $ -0.674$ & $ 0.0434$ & $ -0.717$\\
 \rule[-0.1cm]{0cm}{0.56cm} 
10 & 
$ \approx -0.01157$ & 
$ \approx -0.00645$ & $  0.749$ & $ -0.680$ & $ 0.0344$ & $ -0.715$\\
 \rule[-0.1cm]{0cm}{0.56cm} 
12 & 
$ \approx -0.00771$ & 
$ \approx -0.00409$ & $  0.741$ & $ -0.684$ & $ 0.0285$ & $ -0.713$\\
 \rule[-0.1cm]{0cm}{0.56cm} 
14 & 
$ \approx -0.00551$ & 
$ \approx -0.00282$ & $  0.736$ & $ -0.687$ & $ 0.0243$ & $ -0.712$\\
 \rule[-0.1cm]{0cm}{0.56cm} 
16 & 
$ \approx -0.00414$ & 
$ \approx -0.00206$ & $  0.732$ & $ -0.689$ & $ 0.0212$ & $ -0.711$\\
\hline
\hline
\end{tabular}
\end{center}
\caption[x]{\footnotesize Contributions at each level to coefficients
of terms of order $A^4$}
\label{t:a4}
\end{table}
A more detailed description
of the contribution from each field up to level 6 is given in Appendix
B.
An important feature of these results is the behavior of the sum and
difference of the coefficients $\gamma_1, \gamma_2$,
\begin{equation}
\gamma_\pm = \frac{\gamma_1 \pm \gamma_2}{2} .
\end{equation}
The only way to form a gauge/translation invariant combination of the
two quartic terms in (\ref{eq:quartic-g}) is to have a term of the
form (\ref{eq:f2}), which appears when $\gamma_+ = 0$.  When this is
the case, (\ref{eq:quartic-g}) can be rewritten in the
form
\begin{equation}
-\kappa^2 g^2 \alpha' 
\left(\frac{\gamma_-}{2}   \right)
{\rm Tr}\; \left([A^\mu, A^\nu]^2 \right).
\label{eq:gm}
\end{equation}
The successive approximations to $\gamma_\pm$ at levels $n \leq 16$
are graphed in Figure~\ref{f:gauge}.
From Table 2 and Figure~\ref{f:gauge}, we see that indeed as the
contribution from all levels 
is considered,
\begin{eqnarray}
\gamma_+ & \rightarrow &  0 \nonumber\\
\gamma_- &  \rightarrow &  \approx 0.70
\label{eq:gn}
\end{eqnarray}
At level 16, the $\gamma_+$ term produced by the tachyon has been
cancelled to within 3\%, and the coefficient of $\gamma_-$ has
converged to within approximately 2\% of its apparent asymptotic
value.  Thus, we see that string field theory satisfies the important
consistency check that the effective U(N) vector theory has a
gauge-invariant quartic term for $A^\mu$ in the effective theory
arising when all massive fields are integrated out, even though gauge
invariance is apparently broken when only a subset of the fields are
integrated out.  The results of this calculation seem to have nice
convergence properties, with slower than exponential convergence just
as the terms discussed previously for the tachyon.  It would be nice
to have a better theoretical understanding of the relative rates of
convergence of these terms in the effective action.

It is interesting to compare our results for the coefficient of the
$[A_\mu, A_\nu]^2$ term with the kinetic and cubic terms for the
vector field in the string field theory action
\begin{equation}
\frac{1}{2}  (\partial_\mu A_\nu)^2
-i g \kappa \sqrt{\alpha'} \left(\frac{2^5 \sqrt{2}}{3^3 \sqrt{3}}  \right)
\partial_\mu A_\nu[A^\mu, A^\nu].
\label{eq:a23}
\end{equation}
From these terms, we might expect to find a quartic term of the form
(\ref{eq:gm}) with coefficient
\begin{equation}
 \gamma_-^{?} = \frac{2^{9}}{3^7}  \approx 0.234
\label{eq:gp}
\end{equation}
This result disagrees with (\ref{eq:gn}) by a significant margin,
perhaps by a factor of 3 or $2\sqrt{2}$.  This discrepancy is somewhat
puzzling.  One obvious possibility is that there is an error somewhere
in the calculation.  There are numerous factors of 2 and $ \sqrt{2}$
appearing in the calculation of (\ref{eq:a23}), but most of these
would affect (\ref{eq:gp}) by an overall factor of 2, so it is
difficult to see how the needed factor could arise.  This discrepancy
does not depend on the details of the calculation at high level;
indeed, the result (\ref{eq:gp}) seems to already be ruled out from
the explicit calculations of the quartic term at level 0 and level 2
given in detail above, assuming that the results for $\gamma_2$ from
successive level truncation are monotonically decreasing as they seem
to be.  Another possible explanation of the discrepancy is that the
gauge of the U(1) theory has been fixed in such a way as to introduce
spurious terms either at order $\partial A [A, A]$ or at order $[A,
A]^2$.  We leave a resolution of this puzzle as an outstanding
problem, but in any case it seems likely that this discrepancy can be
resolved without requiring dramatic changes in the picture we have
presented here.

\section{Further directions}

We conclude with a brief discussion of several further issues.  In
subsection \ref{sec:NBI} we discuss the calculation of higher-order
terms in the gauge field effective action, which for the U(N) theory
may give a form of the nonabelian Born-Infeld theory.  In subsection
\ref{sec:condensation} we discuss the extension of this approach to the
supersymmetric theory and a class of configurations in which a direct
connection can be made between descriptions of tachyon condensation in
string theory and in D-brane world-volume gauge theories.

\subsection{Nonabelian Born-Infeld}
\label{sec:NBI}

In supersymmetric string theories, infinite flat D-branes are stable
BPS objects whose fluctuations are described in flat space by a
Born-Infeld action of the form \cite{Tseytlin-vector,Leigh}
\begin{equation}
S  \sim - T_p\int d^{p + 1}\xi \sqrt{-\det (\eta_{\mu \nu} + 
 F_{\mu \nu})}
\label{eq:Born-Infeld}
\end{equation}
where $p$ is the dimension of the D-brane, $T_p$ is the brane tension
and $F_{\mu \nu}$ is the U(1) field on the world volume of
the D-brane.  Expanding this action in $F$ gives
\begin{equation}
S \sim T_p\int \left(-1 -\frac{1}{4}  F_{\mu \nu} F^{\mu \nu}
+ \frac{1}{8}  \left( F_{\mu \nu} F^{\nu \lambda} F_{\lambda \sigma}
F^{\sigma \mu} -\frac{1}{4}   (F_{\mu \nu} F^{\mu \nu})^2 \right)
+ \cdots \right)
\label{eq:Born-Infeld-4}
\end{equation}
A similar action describes the dynamics of a D-brane in the bosonic
theory.  When there are multiple parallel D-branes of the same type,
just as the leading U(1) Yang-Mills theory is extended to a nonabelian
U(N) Yang-Mills theory, the Born-Infeld action should be extended to a
Nonabelian Born-Infeld (NBI) action.  Determining
the exact structure of this
nonabelian action is a long-standing problem in string theory.
Because the field strength components $F_{\mu \nu}$ do not commute,
the ordering of the terms in the nonabelian generalization of the
action (\ref{eq:Born-Infeld}) is not well-defined without further
information.  In the abelian theory, (\ref{eq:Born-Infeld}) is
corrected by terms of higher order in $\alpha'$ containing derivatives
of $F$.  In the nonabelian theory, it is not possible to separate
these correction terms from the ordering ambiguity just mentioned
since $[D_{\mu} D_{\nu}] F_{\lambda \sigma} \sim F_{\mu \nu}
F_{\lambda \sigma}$.  It was proposed by Tseytlin in \cite{Tseytlin}
that the nonabelian action at leading order in $\alpha'$ should be
given by first expanding in $F$ as in the abelian theory, and then
taking a symmetrized trace in which all possible orderings of the
components $F_{\mu \nu}$ are given an equal weighting.  This
prescription amounts to treating all commutators of $F$'s as
corrections of higher order in $\alpha'$.  In the supersymmetric case,
this proposal agrees with calculations which have been performed in
string theory \cite{Gross-Witten,Tseytlin-vector} and M(atrix) theory
\cite{Chepelev-Tseytlin,Dan-Wati-2} for the $F^4$ terms in
(\ref{eq:Born-Infeld-4}), and for the terms linearly coupling the NBI
action to supergravity background fields
\cite{Mark-Wati-4,Mark-Wati-5,Myers-dielectric}.  For a review of
recent work on the nonabelian Born-Infeld action, see
\cite{Tseytlin-review}.

At this point there is no systematic understanding of the $\alpha'$
corrections to the Born-Infeld action, either in the abelian or
nonabelian theories (for some discussion of these terms, see
\cite{Andreev-Tseytlin,Hashimoto-corrections,Okawa-DBI}).  These terms
are not uniquely defined due to the possibility of changing the
apparent form of the effective action through field redefinitions.  A
particularly interesting set of corrections are those which appear for
constant but noncommuting field strengths $F$.  Such terms are
necessary in order to have agreement between the Born-Infeld and
string theory descriptions of certain brane configurations
\cite{Hashimoto-Taylor,Bain}.

In the supersymmetric nonabelian Born-Infeld theory, there are no
corrections to (\ref{eq:Born-Infeld-4}) up to order $F^5$.  In the
bosonic theory, on the other hand, there are corrections containing
commutators of $F$'s at orders $F^3$ \cite{Scherk-Schwarz}, $F^4$
\cite{Tseytlin-vector} and $F^5$ \cite{Kitazawa}.
An interesting application of the level truncation method in string
field theory would be to extend the analysis of this paper past the
quartic terms in the gauge field and to  derive the higher order
terms in the effective action of the nonabelian gauge field on a
bosonic or supersymmetric
D-brane.  If the action is indeed to take the form of the nonabelian
Born-Infeld action, we expect that the terms at order $A^6$ will all
vanish in the supersymmetric theory, while in the bosonic theory we
should find a term of the form
\cite{Scherk-Schwarz}
\[
F_{\mu}^{\; \nu}[F_{\nu \lambda}, F^{\lambda \mu}].
\]
While only four topological types of diagrams contribute to
these terms, this is a somewhat more challenging calculation than the
determination of the quartic terms performed in this paper, since the
cubic vertices between an arbitrary set of 3 fields must be used in
the calculation if we include all diagrams with fields below a fixed
level.  It would be very interesting to see if the perturbative
calculation of \cite{Scherk-Schwarz} can be correctly reproduced from
open bosonic string field theory.
The next interesting question is whether the terms of order
$A^8$ have the characteristic structure ${\rm Tr}\; ([A, A]^4-([A,
A]^2)^2/4)$ of the Born-Infeld action (\ref{eq:Born-Infeld-4}).
According to \cite{Tseytlin-vector}, in the bosonic theory we should
find in addition to these terms a correction of the form
\[
F_{\mu \nu} F_{\lambda \sigma}[F^{\mu \nu}, F^{\lambda \sigma}]
\]
At order $A^{10}$ we again expect a nonzero result containing
commutator terms \cite{Kitazawa}.  Assuming that all these terms can
be calculated and agree with the predictions of perturbative
calculations, the next interesting question comes in at order
$A^{12}$, corresponding to order $F^6$ in the nonabelian Born-Infeld
action.  If it were possible to use level truncation to convincingly
demonstrate the structure of these terms, it would represent a new
piece of information about the nonabelian Born-Infeld action.  While
these calculations would be fairly complicated, they should in
principle be quite tractable with the aid of automated symbolic
manipulation tools, including all fields at least up to level 8 or 10.
Whether a calculation at this level will be sufficient to fix the
structure of the higher order terms in the action is an interesting
open problem for future research.  If these terms can indeed be
calculated in the bosonic theory it may help us to understand the
structure of higher-derivative corrections in this theory.  If this
approach can be extended to the supersymmetric theory, it might give
valuable insight into the structure of the full supersymmetric
nonabelian Born-Infeld theory.

One subtlety which may complicate the interpretation of the higher
order terms in the zero-momentum vector field action is the issue of
gauge invariance.  In particular, since the kinetic terms in the
vector field action are not gauge invariant, there is no reason why
the zero momentum part of the action must preserve gauge invariance.
Since, however, as discussed above the gauge invariance
of the zero-momentum sector is T-dual to translation invariance, we
are fairly confident that all the terms in the action which do not
have momentum dependence should have the appropriate dependence on
$[A_\mu, A_\nu]$ for a Born-Infeld action.  Only further calculations
or a better understanding of the role of the Feynman-Siegel gauge in
the effective vector field theory will resolve this question completely.

It is also worth pointing out in this context that it may be simpler
to approach the Born-Infeld action by including momentum dependence,
so that the $A^{12}$ term of interest, representing a 12-string
interaction, can instead be determined by calculating a 6-string
interaction with 6 powers of momentum of the form $(\partial A)^{6}$.
While this may indeed be an easier approach, it is still necessary to
include a fairly general class of cubic vertices at this order, and
the simplification in diagrams may not make up for the added
complications of including momentum.  It is also more likely that the
momentum-dependent terms of the $F^6$ part of the action will be
difficult to interpret due to the issues of gauge-dependence mentioned
above.  It will be quite interesting, however, to see whether either
of these approaches can give new information about the nonabelian
Born-Infeld action.

\subsection{Tachyon condensation}
\label{sec:condensation}

A fundamental question about the open bosonic string field theory,
which has been the subject of much study since the early days of the
subject, is that of understanding the mechanism of tachyon
condensation.  It is important to understand whether there is a stable
vacuum in the theory which can be found by giving the fields
expectation values, and if such a vacuum exists it is of interest to
understand the dynamics of fluctuations around that true vacuum for
the theory.  There are many other scenarios in which similar questions
arise, including in particular the tachyonic closed bosonic string
theory and brane-antibrane systems in supersymmetric string theories.
In the case of brane-antibrane systems in type II string theory,
we have a fairly good understanding of the nature of the ground
state: a D-brane and anti D-brane of the same dimension should
annihilate into the ground state of the theory.  
An interesting aspect of some brane-antibrane
systems is that they can be described in field theory as well as in
string theory.  In these cases, the mechanism of tachyon condensation
can be seen explicitly in the field theory.  It would be very
interesting to make a more explicit connection between the string
field theory and field theory descriptions of these configurations; we
devote the last part of this paper to sketching part of this
connection.

\junk{At this point, the formalism for the covariant supersymmetric open
string field theory is less well understood than that for the bosonic
theory.  While the framework of Witten \cite{Witten-SFT} can in
principle be extended to the supersymmetric theory
\cite{Witten-SFT-2,Gross-Jevicki-3}, this program is still incomplete.
A very promising alternative formalism is presented in
\cite{Berkovits-general}.  Recently Berkovits showed that for an
unstable D-brane in the NS sector of the non-GSO projected superstring
an analogous calculation to those of Kostelecky/Samuel and
Sen/Zwiebach gives rise at leading order to a tachyon condensate which
cancels 60\% of the brane energy \cite{Berkovits-tachyon}.  We will
not attempt here to confront any of the challenges of constructing a
consistent open superstring field theory, but will describe a simple
configuration which may help to clarify the process of tachyon
condensation in gauge theory and string field theory.}

As mentioned in the previous subsection, the world-volume dynamics of
a D-brane in type II superstring theory is described by a
supersymmetric Born-Infeld type action.  In the low-energy limit, this
reduces to super Yang-Mills theory in flat space.  A parallel brane
and antibrane of the same dimension which come close enough together
develop a tachyonic instability \cite{Banks-Susskind}.  It is believed
\cite{Sen-tachyon} that the brane and antibrane should annihilate in a
process which may be understood as the tachyon and other string fields
developing nonzero expectation values, leading to a vacuum which is
nonperturbative in the original open string field theory.  If the
brane and antibrane are both free of additional structure, it is
believed that the resulting vacuum is the true vacuum of the type II
string theory.

By placing fluxes on one or both of the D-branes, additional charges
are added to the configuration representing lower-dimensional D-branes
\cite{Douglas} which may persist after the initial brane and antibrane
have annihilated.  This picture forms the basis for
Witten's suggestion \cite{Witten-K,Horava-K} that K-theory is the natural
language for describing the topological quantum numbers associated
with configurations of multiple D-branes.  A parallel brane and
antibrane can be interpreted as two branes with opposite orientation,
and without the addition of further charges or fluxes cannot be given
a simple description in terms of super Yang-Mills theory.  After the
addition of further brane charges encoded in fluxes on the brane and
antibrane, however, it is possible in certain circumstances to
describe the brane-antibrane annihilation process completely in the
language of super Yang-Mills theory.

A configuration of this type was discussed in \cite{Hashimoto-Taylor}.
For concreteness we will frame the following discussion in terms of
the example discussed there.  Consider compactifying type IIA string
theory on a 2-torus with sides of length $L_1, L_2$.  Let us wrap two
D2-branes around the torus with the same orientation.  We place $+1$
unit of magnetic flux on one D-brane and $-1$ unit of magnetic flux on
the other.  This corresponds to placing a D0-brane and an anti
D0-brane on the two D2-branes.  Under a pair of T-dualities, this
configuration goes to a D2-brane and an anti D2-brane each with a unit
of D0-brane charge.  This is an example of the type of system
mentioned above with a tachyonic instability leading to a vacuum with
residual D-branes.  Under a single T-duality, the two D2 $\pm$ D0
branes become a pair of intersecting diagonally wrapped D1-branes.

The spectrum of string excitations around this background was found in
the intersecting D1-brane language by Berkooz, Douglas and Leigh
\cite{bdl}.  When the branes are coincident in the directions
transverse to the torus, the spectrum contains a tachyon and an
infinite tower of massive states with level spacing proportional to
the angle $\theta$ between the D1-branes.  The angle $\theta$ is
related to the magnetic flux density $F$ on the branes in the D2-brane
picture through
\[
\tan (\theta/2) = 2 \pi \alpha' F.
\]

A particularly interesting aspect of this configuration is that the
tachyonic instability as well as the mechanism of tachyon condensation
can be completely understood within the framework of gauge theory.  In
our original system with two D2-branes of the same orientation, we can
define a U(2) gauge field with boundary conditions
\begin{eqnarray*}
A_1 (x_1 + L_1, x_2) & = &  e^{2 \pi i (x_2/L_2) \tau_3} 
 A_1 (x_1, x_2)  e^{-2 \pi i (x_2/L_2) \tau_3}\\
A_1 (x_1, x_2 + L_2) & = &    A_1 (x_1, x_2) \\
A_2 (x_1 + L_1, x_2) & = &  e^{2 \pi i (x_2/L_2) \tau_3}  A_2 (x_1, x_2) 
 e^{-2 \pi i (x_2/L_2) \tau_3}
  + \left(\frac{2 \pi}{L_2 }\right) \tau_3  \\
A_2 (x_1, x_2 + L_2) & = &    A_2 (x_1, x_2)
\end{eqnarray*}
Although these boundary conditions look nontrivial, they are
equivalent to trivial boundary conditions through a gauge
transformation.  We now consider the gauge field background
\begin{eqnarray}
A^0_1 & = &  0\label{eq:background}\\
A^0_2 & = &  \frac{2 \pi}{L_1 L_2}  x_1 \tau_3. \nonumber
\end{eqnarray}
This corresponds to placing a unit of flux on the first D2-brane and a
unit of anti flux on the second D2-brane.  The spectrum of excitations
around this background in the Yang-Mills theory was originally studied
by Van Baal \cite{vb1}. (Van Baal worked on $T^4$, but we can reduce
to the $T^2$ case by setting the curvature $F_{34}$ to vanish.)  This
spectrum indeed contains a tachyon, as well as an infinite series of
massive modes with level spacing proportional to $F$.  In the small
$F$ limit the Yang-Mills and string theory spectra coincide; in
\cite{Hashimoto-Taylor} it is shown that under certain circumstances
the symmetrized trace formulation of nonabelian Born-Infeld theory
produces combinatorial relations which transform $F$ into $\theta
\sim\tan^{-1} F$.
  
The gauge field fluctuations around the background
(\ref{eq:background}) can be written in terms of theta functions on
the torus.  The tachyonic modes are given explicitly in
\cite{Hashimoto-Taylor}, and the theta functions for all the other
modes are given by Troost in \cite{Troost-constant}.  In this
configuration, the entire process by which the tachyonic mode
condenses and the entire system of modes acquire expectation values,
leading to a nonperturbative vacuum, can be seen in the language of
Yang-Mills theory.  Both the initial and final configurations have
diagonal gauge fields, so that the energies can be computed using
abelian Born-Infeld.  The gap between initial and final energies in
the Born-Infeld theory is precisely the difference in energies between
a pair of bound D2 $\pm$ D0 brane systems and a pair of D2-branes.

We propose that this system may be an excellent model with which learn
more about the mechanism of tachyon condensation in string field
theory.  The infinite tower of theta function modes in the gauge
theory picture precisely corresponds to the infinite tower of string
modes stretching between the branes in the D1-brane picture.  It
should in principle be possible to go from the known spectrum of
excitations on the string side to an open string field theory.  The
effective action of the set of string fields which correspond to the
Yang-Mills fluctuations should then at quartic order precisely
reproduce the Yang-Mills action in terms of the diagonal and theta
function fluctuations on the D2-brane world-volume.  At higher order,
as discussed in the preceding section, we should see the structure of
the nonabelian Born-Infeld theory which allows the initial unstable
but abelian vacuum to decay into the final stable vacuum.  Some
progress in understanding tachyon condensation in this system from the
string theory side was made by Gava, Narain and Sarmadi \cite{gns},
who showed that when the tachyon condenses to the minimum of the
potential arising when quartic terms computed from string theory are
included, the energy gap is on the same order as the energy
differential between the unstable and stable brane configurations.  So
far, however, this configuration is not completely understood in the
language of string field theory.  Indeed, as mentioned above there are
technical challenges remaining to finding a consistent calculable
formulation of open or closed superstring field theory, but it may be
that the parallel presented here with tachyonic configurations in
gauge theory will be a helpful clue to resolving some of these
difficulties.

\section*{Acknowledgments}

I would like to thank Barton Zwiebach for interesting me in this
subject and for numerous helpful explanations, discussions, comments
and suggestions.  Thanks also to Leonardo Rastelli for helpful
conversations, particularly regarding the bosonic Born-Infeld theory.
This work was supported in part by the A.\ P.\ Sloan Foundation and in
part by the DOE through contract \#DE-FC02-94ER40818.

\newpage
\appendix

\section{Tables of coefficients $N^{rs}_{nm}, X^{rs}_{nm}$}

These are the coefficients appearing in the Witten vertex for the
bosonic string (\ref{eq:interactions}), computed from equations
(\ref{eq:nx}, \ref{eq:n})
\begin{center}
\begin{tabular}{| | c | c | | c | c | c | | }
\hline
\hline
  $n$ &  $m$
 & $N^{11}_{nm}$ & $N^{12}_{nm}$ & $N^{13}_{nm}$
\\
\hline
\hline
1 & 1
 & 
$-5/27$
 & 
$16/27$
 & 
$16/27$
\\ \hline
\hline
1 & 2
 & 
0
 & 
$32\,{\sqrt{3}}/243$
 & 
$-32\,{\sqrt{3}}/243$
\\ \hline
2 & 1
 & 
0
 & 
$-32\,{\sqrt{3}}/243$
 & 
$32\,{\sqrt{3}}/243$
\\ \hline
\hline
1 & 3
 & 
$32/729$
 & 
$-16/729$
 & 
$-16/729$
\\ \hline
2 & 2
 & 
$13/486$
 & 
$-64/243$
 & 
$-64/243$
\\ \hline
3 & 1
 & 
$32/729$
 & 
$-16/729$
 & 
$-16/729$
\\ \hline
\hline
1 & 4
 & 
0
 & 
$-64\,{\sqrt{3}}/2187$
 & 
$64\,{\sqrt{3}}/2187$
\\ \hline
2 & 3
 & 
0
 & 
$-160\,{\sqrt{3}}/2187$
 & 
$160\,{\sqrt{3}}/2187$
\\ \hline
3 & 2
 & 
0
 & 
$160\,{\sqrt{3}}/2187$
 & 
$-160\,{\sqrt{3}}/2187$
\\ \hline
4 & 1
 & 
0
 & 
$64\,{\sqrt{3}}/2187$
 & 
$-64\,{\sqrt{3}}/2187$
\\ \hline
\hline
1 & 5
 & 
$-416/19683$
 & 
$208/19683$
 & 
$208/19683$
\\ \hline
2 & 4
 & 
$-256/19683$
 & 
$128/19683$
 & 
$128/19683$
\\ \hline
3 & 3
 & 
$-893/59049$
 & 
$10288/59049$
 & 
$10288/59049$
\\ \hline
4 & 2
 & 
$-256/19683$
 & 
$128/19683$
 & 
$128/19683$
\\ \hline
5 & 1
 & 
$-416/19683$
 & 
$208/19683$
 & 
$208/19683$
\\ \hline
\hline
1 & 6
 & 
0
 & 
$800\,{\sqrt{3}}/59049$
 & 
$-800\,{\sqrt{3}}/59049$
\\ \hline
2 & 5
 & 
0
 & 
$352\,{\sqrt{3}}/19683$
 & 
$-352\,{\sqrt{3}}/19683$
\\ \hline
3 & 4
 & 
0
 & 
$3136\,{\sqrt{3}}/59049$
 & 
$-3136\,{\sqrt{3}}/59049$
\\ \hline
4 & 3
 & 
0
 & 
$-3136\,{\sqrt{3}}/59049$
 & 
$3136\,{\sqrt{3}}/59049$
\\ \hline
5 & 2
 & 
0
 & 
$-352\,{\sqrt{3}}/19683$
 & 
$352\,{\sqrt{3}}/19683$
\\ \hline
6 & 1
 & 
0
 & 
$-800\,{\sqrt{3}}/59049$
 & 
$800\,{\sqrt{3}}/59049$
\\ \hline
\hline
1 & 7
 & 
$2272/177147$
 & 
$-1136/177147$
 & 
$-1136/177147$
\\ \hline
2 & 6
 & 
$1408/177147$
 & 
$-704/177147$
 & 
$-704/177147$
\\ \hline
3 & 5
 & 
$1504/177147$
 & 
$-752/177147$
 & 
$-752/177147$
\\ \hline
4 & 4
 & 
$5125/708588$
 & 
$-22784/177147$
 & 
$-22784/177147$
\\ \hline
5 & 3
 & 
$1504/177147$
 & 
$-752/177147$
 & 
$-752/177147$
\\ \hline
6 & 2
 & 
$1408/177147$
 & 
$-704/177147$
 & 
$-704/177147$
\\ \hline
7 & 1
 & 
$2272/177147$
 & 
$-1136/177147$
 & 
$-1136/177147$
\\ \hline
\hline
1 & 8
 & 
0
 & 
$-38272\,{\sqrt{3}}/4782969$
 & 
$38272\,{\sqrt{3}}/4782969$
\\ \hline
2 & 7
 & 
0
 & 
$-41312\,{\sqrt{3}}/4782969$
 & 
$41312\,{\sqrt{3}}/4782969$
\\ \hline
3 & 6
 & 
0
 & 
$-203296\,{\sqrt{3}}/14348907$
 & 
$203296\,{\sqrt{3}}/14348907$
\\ \hline
4 & 5
 & 
0
 & 
$-195136\,{\sqrt{3}}/4782969$
 & 
$195136\,{\sqrt{3}}/4782969$
\\ \hline
5 & 4
 & 
0
 & 
$195136\,{\sqrt{3}}/4782969$
 & 
$-195136\,{\sqrt{3}}/4782969$
\\ \hline
6 & 3
 & 
0
 & 
$203296\,{\sqrt{3}}/14348907$
 & 
$-203296\,{\sqrt{3}}/14348907$
\\ \hline
7 & 2
 & 
0
 & 
$41312\,{\sqrt{3}}/4782969$
 & 
$-41312\,{\sqrt{3}}/4782969$
\\ \hline
8 & 1
 & 
0
 & 
$38272\,{\sqrt{3}}/4782969$
 & 
$-38272\,{\sqrt{3}}/4782969$
\\ \hline
\hline
\end{tabular}
\end{center}

\newpage

\begin{center}
\begin{tabular}{| | c | c | | c | c | c | | }
\hline
\hline
  $n$ &  $m$
 & $X^{11}_{nm}$ & $X^{12}_{nm}$ & $X^{13}_{nm}$\\
\hline
\hline
1 & 1
 & 
$11/27$
 & 
$8/27$
 & 
$8/27$
\\ \hline
\hline
1 & 2
 & 
0
 & 
$40\,{\sqrt{3}}/243$
 & 
$-40\,{\sqrt{3}}/243$
\\ \hline
2 & 1
 & 
0
 & 
$-80\,{\sqrt{3}}/243$
 & 
$80\,{\sqrt{3}}/243$
\\ \hline
\hline
1 & 3
 & 
$-80/729$
 & 
$40/729$
 & 
$40/729$
\\ \hline
2 & 2
 & 
$-19/243$
 & 
$-112/243$
 & 
$-112/243$
\\ \hline
3 & 1
 & 
$-80/243$
 & 
$40/243$
 & 
$40/243$
\\ \hline
\hline
1 & 4
 & 
0
 & 
$-104\,{\sqrt{3}}/2187$
 & 
$104\,{\sqrt{3}}/2187$
\\ \hline
2 & 3
 & 
0
 & 
$-304\,{\sqrt{3}}/2187$
 & 
$304\,{\sqrt{3}}/2187$
\\ \hline
3 & 2
 & 
0
 & 
$152\,{\sqrt{3}}/729$
 & 
$-152\,{\sqrt{3}}/729$
\\ \hline
4 & 1
 & 
0
 & 
$416\,{\sqrt{3}}/2187$
 & 
$-416\,{\sqrt{3}}/2187$
\\ \hline
\hline
1 & 5
 & 
$1136/19683$
 & 
$-568/19683$
 & 
$-568/19683$
\\ \hline
2 & 4
 & 
$800/19683$
 & 
$-400/19683$
 & 
$-400/19683$
\\ \hline
3 & 3
 & 
$2099/19683$
 & 
$8792/19683$
 & 
$8792/19683$
\\ \hline
4 & 2
 & 
$1600/19683$
 & 
$-800/19683$
 & 
$-800/19683$
\\ \hline
5 & 1
 & 
$5680/19683$
 & 
$-2840/19683$
 & 
$-2840/19683$
\\ \hline
\hline
1 & 6
 & 
0
 & 
$1528\,{\sqrt{3}}/59049$
 & 
$-1528\,{\sqrt{3}}/59049$
\\ \hline
2 & 5
 & 
0
 & 
$688\,{\sqrt{3}}/19683$
 & 
$-688\,{\sqrt{3}}/19683$
\\ \hline
3 & 4
 & 
0
 & 
$3320\,{\sqrt{3}}/19683$
 & 
$-3320\,{\sqrt{3}}/19683$
\\ \hline
4 & 3
 & 
0
 & 
$-13280\,{\sqrt{3}}/59049$
 & 
$13280\,{\sqrt{3}}/59049$
\\ \hline
5 & 2
 & 
0
 & 
$-1720\,{\sqrt{3}}/19683$
 & 
$1720\,{\sqrt{3}}/19683$
\\ \hline
6 & 1
 & 
0
 & 
$-3056\,{\sqrt{3}}/19683$
 & 
$3056\,{\sqrt{3}}/19683$
\\ \hline
\hline
1 & 7
 & 
$-6640/177147$
 & 
$3320/177147$
 & 
$3320/177147$
\\ \hline
2 & 6
 & 
$-4640/177147$
 & 
$2320/177147$
 & 
$2320/177147$
\\ \hline
3 & 5
 & 
$-3568/59049$
 & 
$1784/59049$
 & 
$1784/59049$
\\ \hline
4 & 4
 & 
$-8251/177147$
 & 
$-84448/177147$
 & 
$-84448/177147$
\\ \hline
5 & 3
 & 
$-17840/177147$
 & 
$8920/177147$
 & 
$8920/177147$
\\ \hline
6 & 2
 & 
$-4640/59049$
 & 
$2320/59049$
 & 
$2320/59049$
\\ \hline
7 & 1
 & 
$-46480/177147$
 & 
$23240/177147$
 & 
$23240/177147$
\\ \hline
\hline
1 & 8
 & 
0
 & 
$-82040\,{\sqrt{3}}/4782969$
 & 
$82040\,{\sqrt{3}}/4782969$
\\ \hline
2 & 7
 & 
0
 & 
$-86288\,{\sqrt{3}}/4782969$
 & 
$86288\,{\sqrt{3}}/4782969$
\\ \hline
3 & 6
 & 
0
 & 
$-239720\,{\sqrt{3}}/4782969$
 & 
$239720\,{\sqrt{3}}/4782969$
\\ \hline
4 & 5
 & 
0
 & 
$-763424\,{\sqrt{3}}/4782969$
 & 
$763424\,{\sqrt{3}}/4782969$
\\ \hline
5 & 4
 & 
0
 & 
$954280\,{\sqrt{3}}/4782969$
 & 
$-954280\,{\sqrt{3}}/4782969$
\\ \hline
6 & 3
 & 
0
 & 
$479440\,{\sqrt{3}}/4782969$
 & 
$-479440\,{\sqrt{3}}/4782969$
\\ \hline
7 & 2
 & 
0
 & 
$302008\,{\sqrt{3}}/4782969$
 & 
$-302008\,{\sqrt{3}}/4782969$
\\ \hline
8 & 1
 & 
0
 & 
$656320\,{\sqrt{3}}/4782969$
 & 
$-656320\,{\sqrt{3}}/4782969$
\\ \hline
\hline
\end{tabular}
\end{center}
\vspace{0.2in}

\noindent
The coefficients
$N^{rs}_{nm}, X^{rs}_{nm}$ have a cyclic symmetry under $r \rightarrow
(r\;{\rm mod} \; 3) + 1 $, $s \rightarrow (s\;{\rm mod} \; 3) + 1 $, so we
only give the coefficients for $r = 1$.

\section{Some details of the calculations up to level  6}

This table describes the contributions of each state to the quartic
terms in the effective tachyon and gauge field potential.  Note that,
like  $B_{\mu \nu}$, many fields can be contracted in multiple ways
with a pair of scalar or vector fields.  The contribution calculated
includes interactions from all such contractions.

\begin{center}
\begin{tabular}{| | r | r  | r |  r | | }
\hline
\hline
state & $\hat{\gamma}$& $\hat{\gamma}_1 $& $\hat{\gamma}_2$\\
\hline
\hline
$
| 0 \rangle $
 \rule[-0.1cm]{0cm}{0.56cm} & 
& 
$\frac{
2^{7}
}{
3^{4}
}$
$ \approx  1.5802$
& 
0
\\
$
b_{-1}
c_{-1}
| 0 \rangle $
 \rule[-0.1cm]{0cm}{0.56cm} & 
$\frac{
11^{2}
}{
2^{}
\cdot
3^{4}
}$
$ \approx  0.7469$
& 
$\frac{
2^{7}
\cdot
11^{2}
}{
3^{10}
}$
$ \approx  0.2623$
& 
0
\\
$
\alpha_{-1}^{\mu}
\alpha_{-1}^{\nu}
| 0 \rangle $
 \rule[-0.1cm]{0cm}{0.56cm} & 
$-\frac{
5^{2}
\cdot
13^{}
}{
2^{}
\cdot
3^{4}
}$
$ \approx -2.0062$
& 
$-\frac{
2^{7}
\cdot
421^{}
}{
3^{10}
}$
$ \approx -0.9126$
& 
$-\frac{
2^{15}
}{
3^{10}
}$
$ \approx -0.5549$
\\
$
b_{-1}
c_{-3}
| 0 \rangle $
 \rule[-0.1cm]{0cm}{0.56cm} & 
$\frac{
2^{7}
\cdot
5^{2}
}{
3^{10}
}$
$ \approx  0.0542$
& 
$\frac{
2^{15}
\cdot
5^{2}
}{
3^{16}
}$
$ \approx  0.0190$
& 
0
\\
$
b_{-2}
c_{-2}
| 0 \rangle $
 \rule[-0.1cm]{0cm}{0.56cm} & 
$\frac{
19^{2}
}{
2^{}
\cdot
3^{9}
}$
$ \approx  0.0092$
& 
$\frac{
2^{7}
\cdot
19^{2}
}{
3^{15}
}$
$ \approx  0.0032$
& 
0
\\
$
b_{-3}
c_{-1}
| 0 \rangle $
 \rule[-0.1cm]{0cm}{0.56cm} & 
$\frac{
2^{7}
\cdot
5^{2}
}{
3^{10}
}$
$ \approx  0.0542$
& 
$\frac{
2^{15}
\cdot
5^{2}
}{
3^{16}
}$
$ \approx  0.0190$
& 
0
\\
$
\alpha_{-3}^{\mu}
\alpha_{-1}^{\nu}
| 0 \rangle $
 \rule[-0.1cm]{0cm}{0.56cm} & 
$-\frac{
2^{10}
\cdot
13^{}
}{
3^{10}
}$
$ \approx -0.2254$
& 
$-\frac{
2^{16}
\cdot
7^{2}
}{
3^{16}
}$
$ \approx -0.0746$
& 
$-\frac{
2^{16}
}{
3^{16}
}$
$ \approx -0.0015$
\\
$
\alpha_{-2}^{\mu}
\alpha_{-2}^{\nu}
| 0 \rangle $
 \rule[-0.1cm]{0cm}{0.56cm} & 
$-\frac{
13^{3}
}{
2^{}
\cdot
3^{9}
}$
$ \approx -0.0558$
& 
$-\frac{
2^{7}
\cdot
7^{}
\cdot
3739^{}
}{
3^{17}
}$
$ \approx -0.0259$
& 
$-\frac{
2^{21}
}{
3^{17}
}$
$ \approx -0.0162$
\\
$
\alpha_{-1}^{\mu}
\alpha_{-1}^{\nu}
b_{-1}
c_{-1}
| 0 \rangle $
 \rule[-0.1cm]{0cm}{0.56cm} & 
$\frac{
5^{2}
\cdot
11^{2}
\cdot
13^{}
}{
2^{}
\cdot
3^{11}
}$
$ \approx  0.1110$
& 
$\frac{
2^{7}
\cdot
11^{2}
\cdot
421^{}
}{
3^{17}
}$
$ \approx  0.0505$
& 
$\frac{
2^{15}
\cdot
11^{2}
}{
3^{17}
}$
$ \approx  0.0307$
\\
$
\alpha_{-1}^{\mu}
\alpha_{-1}^{\nu}
\alpha_{-1}^{\lambda}
\alpha_{-1}^{\rho}
| 0 \rangle $
 \rule[-0.1cm]{0cm}{0.56cm} & 
$-\frac{
5^{4}
\cdot
13^{}
}{
2^{3}
\cdot
3^{10}
}$
$ \approx -0.0172$
& 
$-\frac{
2^{7}
\cdot
5^{2}
\cdot
17^{}
}{
3^{12}
}$
$ \approx -0.1024$
& 
$-\frac{
2^{15}
\cdot
5^{3}
}{
3^{16}
}$
$ \approx -0.0952$
\\
$
b_{-1}
c_{-5}
| 0 \rangle $
 \rule[-0.1cm]{0cm}{0.56cm} & 
$ \approx  0.0150$
& 
$ \approx  0.0053$
& 
0
\\
$
b_{-2}
c_{-4}
| 0 \rangle $
 \rule[-0.1cm]{0cm}{0.56cm} & 
$ \approx  0.0030$
& 
$ \approx  0.0010$
& 
0
\\
$
b_{-3}
c_{-3}
| 0 \rangle $
 \rule[-0.1cm]{0cm}{0.56cm} & 
$ \approx  0.0102$
& 
$ \approx  0.0036$
& 
0
\\
$
b_{-4}
c_{-2}
| 0 \rangle $
 \rule[-0.1cm]{0cm}{0.56cm} & 
$ \approx  0.0030$
& 
$ \approx  0.0010$
& 
0
\\
$
b_{-5}
c_{-1}
| 0 \rangle $
 \rule[-0.1cm]{0cm}{0.56cm} & 
$ \approx  0.0150$
& 
$ \approx  0.0053$
& 
0
\\
$
b_{-2}
b_{-1}
c_{-2}
c_{-1}
| 0 \rangle $
 \rule[-0.1cm]{0cm}{0.56cm} & 
$ \approx -0.0009$
& 
$ \approx -0.0003$
& 
0
\\
$
\alpha_{-5}^{\mu}
\alpha_{-1}^{\nu}
| 0 \rangle $
 \rule[-0.1cm]{0cm}{0.56cm} & 
$ \approx -0.0523$
& 
$ \approx -0.0173$
& 
$ \approx -0.0004$
\\
$
\alpha_{-4}^{\mu}
\alpha_{-2}^{\nu}
| 0 \rangle $
 \rule[-0.1cm]{0cm}{0.56cm} & 
$ \approx -0.0317$
& 
$ \approx -0.0105$
& 
$ \approx -0.0019$
\\
$
\alpha_{-3}^{\mu}
\alpha_{-3}^{\nu}
| 0 \rangle $
 \rule[-0.1cm]{0cm}{0.56cm} & 
$ \approx -0.0241$
& 
$ \approx -0.0084$
& 
$ \approx -0.0000$
\\
$
\alpha_{-1}^{\mu}
\alpha_{-1}^{\nu}
b_{-1}
c_{-3}
| 0 \rangle $
 \rule[-0.1cm]{0cm}{0.56cm} & 
$ \approx  0.0145$
& 
$ \approx  0.0066$
& 
$ \approx  0.0040$
\\
$
\alpha_{-1}^{\mu}
\alpha_{-1}^{\nu}
b_{-2}
c_{-2}
| 0 \rangle $
 \rule[-0.1cm]{0cm}{0.56cm} & 
$ \approx  0.0025$
& 
$ \approx  0.0011$
& 
$ \approx  0.0007$
\\
$
\alpha_{-1}^{\mu}
\alpha_{-1}^{\nu}
b_{-3}
c_{-1}
| 0 \rangle $
 \rule[-0.1cm]{0cm}{0.56cm} & 
$ \approx  0.0145$
& 
$ \approx  0.0066$
& 
$ \approx  0.0040$
\\
$
\alpha_{-2}^{\mu}
\alpha_{-1}^{\nu}
b_{-1}
c_{-2}
| 0 \rangle $
 \rule[-0.1cm]{0cm}{0.56cm} & 
0
& 
0
& 
0
\\
$
\alpha_{-2}^{\mu}
\alpha_{-1}^{\nu}
b_{-2}
c_{-1}
| 0 \rangle $
 \rule[-0.1cm]{0cm}{0.56cm} & 
0
& 
0
& 
0
\\
$
\alpha_{-3}^{\mu}
\alpha_{-1}^{\nu}
b_{-1}
c_{-1}
| 0 \rangle $
 \rule[-0.1cm]{0cm}{0.56cm} & 
$ \approx  0.0225$
& 
$ \approx  0.0074$
& 
$ \approx  0.0002$
\\
$
\alpha_{-2}^{\mu}
\alpha_{-2}^{\nu}
b_{-1}
c_{-1}
| 0 \rangle $
 \rule[-0.1cm]{0cm}{0.56cm} & 
$ \approx  0.0056$
& 
$ \approx  0.0026$
& 
$ \approx  0.0016$
\\
$
\alpha_{-3}^{\mu}
\alpha_{-1}^{\nu}
\alpha_{-1}^{\lambda}
\alpha_{-1}^{\rho}
| 0 \rangle $
 \rule[-0.1cm]{0cm}{0.56cm} & 
$ \approx -0.0070$
& 
$ \approx -0.0290$
& 
$ \approx -0.0188$
\\
$
\alpha_{-2}^{\mu}
\alpha_{-2}^{\nu}
\alpha_{-1}^{\lambda}
\alpha_{-1}^{\rho}
| 0 \rangle $
 \rule[-0.1cm]{0cm}{0.56cm} & 
$ \approx -0.0006$
& 
$ \approx -0.0091$
& 
$ \approx -0.0085$
\\
$
\alpha_{-1}^{\mu}
\alpha_{-1}^{\nu}
\alpha_{-1}^{\lambda}
\alpha_{-1}^{\rho}
b_{-1}
c_{-1}
| 0 \rangle $
 \rule[-0.1cm]{0cm}{0.56cm} & 
$ \approx  0.0017$
& 
$ \approx  0.0102$
& 
$ \approx  0.0095$
\\
$
\alpha_{-1}^{\mu}
\alpha_{-1}^{\nu}
\alpha_{-1}^{\lambda}
\alpha_{-1}^{\rho}
\alpha_{-1}^{\sigma}
\alpha_{-1}^{\tau}
| 0 \rangle $
 \rule[-0.1cm]{0cm}{0.56cm} & 
$ \approx -0.0003$
& 
$ \approx -0.0149$
& 
$ \approx -0.0157$
\\
 \hline
\hline
\end{tabular}
\end{center}

\bibliographystyle{plain}

\end{document}